\documentclass[10pt]{article}

\usepackage[T1]{fontenc}
\usepackage[utf8]{inputenc}
\usepackage[english]{babel}
\usepackage{amsmath,latexsym,amsfonts,amsthm,bm,mathtools}
\usepackage{graphicx}
\usepackage[top=1.5in, bottom=1.5in, left=1.25in, right=1.25in]{geometry}
\usepackage{booktabs,url,tikz}
\usepackage[makeroom]{cancel}
\usepackage{dsfont}
\usepackage{relsize}
\usepackage{multirow}

\usepackage{subfig}
\usepackage{multirow}
\usepackage{rotating}
\usepackage{caption}

\usepackage{pdfpages}

\title{A novel CFA\texttt{+}EFA model to detect aberrant respondents}
\author{Niccol\`o Cao$^{1\ast}$, Livio Finos$^{2}$, Luigi Lombardi$^{3}$, Antonio Calcagn\`i$^2$ \\\\
		\footnotesize{\sl $^{1}$ University of Bologna, \sl $^{2}$ University of Padova, \sl $^3$ University of Trento}\\
		\footnotesize{$\ast$ E-mail: niccolo.cao@unibo.it}
	}

\date{}

\begin{document}

\maketitle

\begin{abstract}
Aberrant respondents are common but yet extremely detrimental to the quality of social surveys or questionnaires. Recently, factor mixture models have been employed to identify individuals providing deceptive or careless responses. We propose a comprehensive factor mixture model for {continuous outcomes} that combines confirmatory and exploratory factor models to {classify both the non-aberrant and aberrant respondents}. The flexibility of the proposed {classification model} allows for the identification of two of the most common aberrant response styles, namely faking and careless responding. We validated our approach by means of two simulations and two case studies. The results indicate the effectiveness of the proposed model in dealing with aberrant responses in social and behavioural surveys.
\end{abstract}

\vspace{2cm}

\section{Introduction}\label{sec:1}

In behavioural and social sciences, surveys and questionnaires serve as primary tools for measuring latent variables, such as psychological or social constructs \cite{brown2015cfa}. To ensure the measurement properties of a questionnaire, confirmatory factor analysis (CFA)  is widely employed to verify whether observed variables (i.e., items of a questionnaire) are valid indicators of the latent variables of interest \cite{brown2015cfa}. 
In most CFA applications, a simplified assumption is the \textit{sample homogeneity} \cite{lubke2006distinguishing}, which presupposes homogeneity at both the population and sample levels simultaneously. At the population level, the observed sample is considered as drawn from a single population (i.e., absence of sub-populations), while, at the sample level, the sample is assumed to consist of mutually exchangeable  sampling units \cite{ZUMBO200645}.  However, in applied contexts, this assumption is seldom fulfilled and the  observed samples are realistically composed of two or even more unknown subgroups of respondents, which may originate in the population level and/or in the sample level. This phenomenon is termed \textit{sample heterogeneity} \cite{sawatzky2009sample}. Sample heterogeneity affecting the population level is referred to as \textit{pre-sampling} heterogeneity, which occurs when the population is divided into multiple and unknown sub-populations  \cite{muthen1989latent}. In such cases, the observed sample can be represented as a mixture of an unknown number of subsamples, each possibly related to a distinct CFA model \cite{yung1997finite}. To address \textit{pre-sampling} heterogeneity, factor mixture models (FMMs), are among the most used methods  \cite{lubke2005investigating}. FMMs simultaneously identify homogeneous subsamples from an  heterogeneous observed sample while estimating the factor model parameters for each discovered subsample \cite{lubke2005investigating}. 
Conversely, sample heterogeneity at the sample level is named \textit{post-sampling} heterogeneity, which typically arises from the interaction between respondents and the context of test administration (e.g., high-stakes or low-stakes situations; \cite{tziner2005contextual}). In this case, heterogeneity emerges from individuals who systematically bias their responses by adopting aberrant response processes \cite{groves2011survey}, such as faking or  careless responding \cite{arthur2021lazy}.
As in the case of \textit{pre-sampling} heterogeneity (see \cite{becker2013discovering}), post-sampling heterogeneity could cause erroneous CFA results, biased parameter estimates, and threats to measurement validity (e.g., see \cite{donovan2014impact,kam2015careless}).

Recently, some researchers suggested the use of theoretically-based psychometric models to enhance statistical methods against \textit{post-sampling} heterogeneity \cite{plieninger2018new,merhof2023dynamic}. For instance, Item Response Theory (IRT) mixture models have been recently employed {to combine two} IRT models, one modelling the honest/careful response style and the other one representing the faking or carelessness behaviour (e.g., see \cite{ulitzsch2022response,frick2022modeling}). Although theoretically well-founded, this approach might not be flexible enough to address \textit{post-sampling} heterogeneity from a more data-driven perspective. An alternative solution is to rely on the analysis of covariance matrix as in CFA/EFA methods. Indeed, numerous studies have shown that \textit{post-sampling} heterogeneity generally implies disruptions (e.g., deflation or inflation) of the observed inter-item correlations among indicators of the latent variables of interest \cite{kam2019careless,ward2023dealing,ellingson1999social,griffin2004applicants,lombardi2015sgr}.

Based on this evidence, FMMs have been recently adopted as they offer a suitable framework to  model directly the observed inter-item correlation matrix, providing a more data-driven approach. There have been several attempts to generalize FMMs to deal with post-sampling heterogeneity. For instance, \cite{leite2010detecting} included an additional latent variable for the social-desirability-bias to detect faking, \cite{ziegler2015nature} used a mixture of factor analysis (FA) to differentiate between extreme and slight fakers, \cite{kam2023constrained} constrained the signs of the factor loadings to differentiate the careful and the careless latent components. However, all of these methods essentially depend on external indicators of aberrant behaviours (e.g., covariates or reverse-keyed items) more than modelling the biased subsample's latent structure itself. Indeed, these authors employed a modelling strategy allowing the aberrant latent structure closely to resemble the unbiased latent structure, with only few adjustments (e.g., opposite signs of the factor loadings for the negatively worded items). 

In this article, we present a novel mixture of factor analysers to detect vectors of aberrant responses as those typically encountered in social surveys and questionnaires. {We propose an unsupervised classification model serving as a general model-based screening tool to identify various types of aberrant respondents}. In the spirit of \cite{banfield1993model}, this aim is achieved by specifying a theoretically sound model for the biased subsample, which is plugged into the mixture as a \textit{noise component}. Recent instances of this approach appeared in the family of clusterwise regression models (e.g., see \cite{perrone2023seemingly,punzo2016parsimonious}. In particular, we combined a standard CFA component for the unbiased subsample and an exploratory factor analysis (EFA) for the biased subsample, defining a CFA\texttt{+}EFA mixture model {for continuous outcomes}. {The idea is that, by specifying the target population's latent structure  in the CFA component,  individuals whose responses do not align with the CFA due to aberrant responding would be captured by the EFA component. This is mainly due to the fact that the constrained latent structure imposed by the CFA would leave the divergent patterns of inter-item correlations to be gathered by the unconstrained factorial structure of the EFA model. The  number of EFA latent factors is selected in order to  maximize the absorption of covariation due to various aberrant response biases through a data-driven procedure. Thus, the mixture latent variable assigns respondents based on this specification of the two submodels, without requiring additional information nor in the mixture parameter neither in the submodels.} 
This allows for a broader applicability of the proposed method. Indeed, in empirical research, reliable external indicators of aberrant responding might be unavailable and assumptions on the sign of factor loadings  for reverse-keyed items has currently a theoretical motivation only in the context of careless responding. 

The remainder of the article is organized as follows. Section \ref{sec:2} provides an overview of the psychometric literature on aberrant response behaviours along with the methods used to detect them. Section \ref{sec:3} presents the CFA\texttt{+}EFA model and the parameters estimation procedure. Section \ref{sec:4} reports the results of two simulation studies validating our model as a classification method. Section \ref{sec:5} describes an application  of the model to two case studies involving ratings data affected by faking and careless responding. Finally, Section \ref{sec:6} concludes the article providing final remarks and suggesting future directions. Technical details as well as additional results are available as supplementary material to this article. Note that the algorithms and scripts used throughout the article are available to download at \url{https://github.com/niccolocao/CFAmixEFA}.

\section{Behavioural sources of post-sampling heterogeneity}\label{sec:2}

As established by the psychometric literature,  a number of aberrant response behaviours can lead to \textit{post-sampling} heterogeneity, including sleeping, guessing, and plodding behaviour \cite{wright1979best}, alignment errors \cite{hulin1983item}, cheating \cite{emons2003person}, creative responding \cite{karabatsos2003comparing}, faking \cite{maccann2011faking}, and careless responding \cite{ward2023dealing}. Among them, faking and careless responding have been widely recognized as major  sources of \textit{post-sampling} heterogeneity \cite{arthur2021lazy}. In what follows, we provide a brief qualitative description of those response styles, which constitute the focus of the method presented in Section \ref{sec:3}.\\

\noindent \textbf{Faking behaviour}. Faking is usually defined as a goal-directed distortion of the response process, entailing the  manipulation of an individual's self-presentation in relation to the test administration context \cite{maccann2011faking}. In particular, a faker (i.e., respondent who adopt faking behaviour) may alter his/her genuine self-presentation by exaggerating positive qualities, possibly associated with a complementary  denial of negative qualities (i.e., faking "good"), and/or by exaggerating negative qualities (i.e., faking "bad") \cite{maccann2011faking}. For instance, faking good is prevalent among job applicants involved in selection procedures \cite{arthur2021lazy}, while faking bad is often observed in cases such as fabrication of clinical symptoms for a secondary gain \cite{hall2011plaintiffs}. The prevalence of faking good and bad varies among different settings. Considering job applicants, approximately in the range 20\% and 40\% of the sample engages in faking good \cite{griffith2011new}. Similarly, in the forensic settings, the prevalence of faking bad has been reported to be approximately  $17.4\%$ \cite{rogers1998comparison} and $40\%$ \cite{mittenberg2002base}.
Literature also suggests that fakers adopt different styles or patterns to fake, such as extreme faking and slight faking \cite{ziegler2015nature}. In particular, the extreme faking response style involves a significant positive shift in the value of the uncorrupted response to an item, while the slight faking  response style implies a moderate positive shift \cite{lombardi2015sgr}. Many studies have highlighted the detrimental effects of faking on inter-items correlations (e.g., \cite{ellingson1999social,lee2019investigating,lee2017application}). In particular, faking can deteriorate a  multidimensional latent structure, causing it to collapse into a single comprehensive faking factor \cite{ellingson1999social,schmit1993big}. Consistently with these findings, some researchers have recently identified more stereotypical responses within faking samples, which resulted in inflated inter-items correlations \cite{schermer2019general2} and increased covariance among the latent factors \cite{schermer2019general1}, reducing the separation between these factors. Interestingly, \cite{lee2019investigating} conducted a simulation study to explore the bias induced by faking on the CFA model estimation, indicating less bias when fitting a CFA model with an additional factor loaded onto all the items. These considerations suggest that a one-factor latent structure for a sample of fakers may be a reasonable hypothesis. \\

\noindent \textbf{Careless responding}. Careless responding refers to a behaviour characterized by random, inattentive, or inconsistent responses, which reflects a diminished consideration of the item content  in the rating process, thus leading to inaccurate responses \cite{ward2023dealing}. Researchers suggested that careless responding may affect most of the survey datasets \cite{ward2023dealing}. The rate of careless respondents can vary widely, ranging from 3.5\% \cite{johnson2005ascertaining} to as high as 46\% \cite{oppenheimer2009instructional} of the sample, with a modal rate  around 15\% \cite{jones2022careless,curran2016methods}. However, even rates  as low as 5\% could significantly compromise the psychometric validity of a questionnaire in terms of inter-items correlations \cite{crede2010random}.
In empirical contexts, careless responding can manifest as either random or systematic behaviour \cite{hong2020methods}: (i) \textit{random responding} involves endorsing one of the response categories for each item in an (almost) random fashion \cite{meade2012identifying}; (ii) \textit{straight-lining}, on the other hand, is a well-studied form of systematic careless responding, where the same configuration of response categories is selected for all items \cite{johnson2005ascertaining}.
These two usual manifestations of careless responding generally result in opposite effects on inter-item correlations. Random responding typically induces weaker correlations, as the responses would asymptotically tend to be uniformly distributed among the response categories \cite{desimone2018differential}. In contrast, straight-lining tends to produce stronger correlations, as the responses remain relatively invariant across the items (e.g.: the selection of a single response category for all the items, \cite{desimone2018differential}). However, in empirical contexts, respondents may vary their level of engagement and shift among different forms of careless responding  \cite{ward2023dealing}.  For instance, \cite{desimone2018differential} demonstrated that if a respondent adopts straight-lining for a small portion of  items, correlations will decrease. Still, if this behaviour extends to a greater number of items, correlations will increase.
The existing literature has shown that the impact of careless responding on inter-item correlations is often unpredictable \cite{huang2015insufficient,ward2023dealing}. 

\section{The CFA\texttt{+}EFA model}\label{sec:3}

Let $\boldsymbol{Y}\in\mathbb{R}^{p}$ denotes the $p\times1$ random vector of the observed indicators. For the CFA component, the random vector $\boldsymbol{Y}$ can be reduced to $q<p$ latent factors being modelled as $\boldsymbol{\eta}\sim \mathcal{N}_q(\boldsymbol{\mu},\boldsymbol{\Phi})$, where $\boldsymbol{\mu}$ and $\boldsymbol{\Phi}$ are the $q$-dimensional vector of the factor means and the $q\times q$ correlation matrix of the latent factors, respectively. In addition, we assumed that the $p\times1$ vector of model errors takes the form $\boldsymbol{\delta}\sim N_p(\boldsymbol{0}_p,\boldsymbol{\Theta_\delta})$, where $\boldsymbol{\Theta_\delta}$ is the $p\times p$ positive definite and diagonal covariance matrix. For the EFA component, the  vector $\boldsymbol{Y}$ can be reduced to $K<p$  latent factors being modelled as $\boldsymbol{\xi}\sim \mathcal{N}_K(\boldsymbol{\nu},\boldsymbol{I}_K)$, where $\boldsymbol{\nu}$ and $\boldsymbol{I}_K$ are the $K$-dimensional vector of the factor means and the $K\times K$ identity covariance matrix of the latent factors, respectively. We considered that the $p\times1$ vector of model errors is defined as $\boldsymbol{\epsilon}\sim N_p(\boldsymbol{0}_p,\boldsymbol{\Psi_\epsilon})$, where  $\boldsymbol{\Psi_\epsilon}$ indicates the $p\times p$ positive definite and diagonal covariance matrix.  
{Let $Z_i\sim\mathcal{B}ern(\pi)$  with $i=1,\dots,n$ be the random vector of the Bernoulli mixture latent variable, which selects the $i$-th observation $\boldsymbol{Y}_i$ as generated from one of the two FA models, and $\pi$ the mixture parameter}. 
Then, the CFA and EFA density components are defined as follows:
\begin{equation*}
	\begin{split}
		\boldsymbol{Y}_i\mid Z_i=1&\sim N_p(\boldsymbol{\Lambda}_1\boldsymbol{\mu},\boldsymbol{\Lambda}_1\boldsymbol{\Phi}\boldsymbol{\Lambda}_1^{T}+\boldsymbol{\Theta_\delta})
	\end{split}
	\quad\quad
	\begin{split}
		\boldsymbol{Y}_i\mid Z_i=0&\sim N_K(\boldsymbol{\Lambda}_2\boldsymbol{\nu},\boldsymbol{\Lambda}_2\boldsymbol{\Lambda}_2^{T}+\boldsymbol{\Psi_\epsilon})
	\end{split}
\end{equation*}
\noindent where  $\boldsymbol{\Lambda}_1$ denotes the $p\times q$ factor loadings matrix which contains some loadings fixed to zero (according to the researcher's hypotheses), and $\boldsymbol{\Lambda}_2$ denotes the $p\times K$ factor loadings matrix (with no fix-to-zero entries). {Furthermore, we assumed that the latent variable $Z_i$ can also be modelled by a logit model, adding covariates to the mixture parameter $\pi$}:
\begin{equation*}
	\pi_i=f(z_i=1;\boldsymbol{\beta}_i)=\frac{\exp(\boldsymbol{x}_i\boldsymbol{\beta})}{1+\exp(\boldsymbol{x}_i\boldsymbol{\beta})}
\end{equation*}
where $\boldsymbol{x}_i$ is the $1\times (C+1)$ vector of $C$ fixed covariates for the $i$-th observation, and $\boldsymbol{\beta}$ is the $(C+1)\times1$ vector of parameter.  The p.d.f. of the realization $\boldsymbol{y}_i$ is then:
\begin{equation}
	f(\boldsymbol{y}_i;\boldsymbol{\Omega})=\pi_i(\boldsymbol{\beta})\mathcal{N}_p(\boldsymbol{y}_i;\boldsymbol{\Lambda}_1\boldsymbol{\mu},\boldsymbol{\Lambda}_1\boldsymbol{\Phi}\boldsymbol{\Lambda}_1^{T}+\boldsymbol{\Theta_\delta})+\left(1-\pi_i(\boldsymbol{\beta})\right)\mathcal{N}_K(\boldsymbol{y}_i;\boldsymbol{\Lambda}_2\boldsymbol{\nu},\boldsymbol{\Lambda}_2\boldsymbol{\Lambda}_2^{T}+\boldsymbol{\Psi_\epsilon})
\end{equation}

\noindent with $\boldsymbol{\Omega}=\{\boldsymbol{\Lambda}_1,\boldsymbol{\mu},\boldsymbol{\Phi},\boldsymbol{\Theta_\delta},\boldsymbol{\Lambda}_2,\boldsymbol{\nu},\boldsymbol{\Psi_\epsilon},\boldsymbol{\beta}\}$ being the vector of model parameters. In general, the CFA\texttt{+}EFA model requires the following constrains to ensure identifiability \cite{brown2015cfa}: (i) $\text{diag}(\boldsymbol{\Phi})=\mathbf 1_q$; (ii) $\boldsymbol{\Theta_\delta}$ is diagonal; (iii) $\boldsymbol{\Psi_\epsilon}$ is diagonal. Given its nature, the CFA\texttt{+}EFA model can be considered as a classification model for identifying aberrant respondents. In this case, the computation of the posterior classification probabilities for the $i$-th observation is performed via the Bayes's theorem:
\begin{align}\label{postclass}
	f(z_{i}=w\mid \boldsymbol{y}_i;\boldsymbol{\Omega})&=\frac{f(z_i=w;\boldsymbol{\beta})f(\boldsymbol{y}_i\mid z_i=w;\boldsymbol{\Omega})}{\pi_if(\boldsymbol{y}_i\mid z_i=1;\boldsymbol{\Omega})+\left(1-\pi_i\right)f\left(\boldsymbol{y}_i\mid z_{i}=0;\boldsymbol{\Omega}\right)},\quad w=0,1
\end{align}

\noindent which represents the probability that $\boldsymbol{y}_i$ belongs to one of the mixture components conditional on the observed sample. To determine the individuals' component assignments, we consider the highest posterior probability estimated for the $i$-th observation between the two components. \\
Despite the simplicity of the CFA\texttt{+}EFA formulation, some advantages need to be highlighted with regard to the EFA component. In general, the latter offers a simple and quite general solution because of the flexibility offered by the unconstrained matrix $\boldsymbol{\Lambda}_2$ \cite{de2017mixture}. From a theoretical standpoint, the potential of the EFA component to represent aberrant respondents becomes evident if one consider faking and  careless responding scenarios. In the first case, research have shown that faking behaviour can alter inter-item correlations as a phenomenon mainly due to the relevance of item content for the respondent's self-presentation objectives \cite{donovan2014impact}. This can lead to a non-random and distinctive latent structure within the faking subsample \cite{bensch2019nature}. Instead, in the second case, careless responding may result in an unpredictable pattern of correlations among indicators \cite{desimone2018differential,arthur2021lazy}, which may result in spurious latent factors unrelated to those of careful respondents. In both instances, the inflated or deflated inter-item correlations stemming from aberrant responding can be suitably captured by the additional and independent latent structure induced by the EFA submodel. It should be stressed that the ability of EFA to recover the aberrant or biased subsample's latent structure is intimately connected with the appropriate selection of $K$, namely the number of latent components in the EFA submodel. In the general framework of FMMs, researchers have used both information (IC) - such as Akaike Information Criterion (AIC), consistent Akaike Information Criterion (CAIC),  Bayesian Information Criterion (BIC), sample-adjusted BIC (ssBIC) - and classification criteria (CC) - such as Classification Likelihood Information Criterion (CLC), Integrated Classification Likelihood-BIC (ICL-BIC), entropy (H) -  to select the best $\hat K$. Despite their differences, both ICs and CCs are usually consistent in selecting the correct number of latent components $K$, although they can be differently sensitive to sample size \cite{henson2007detecting,mclachlan2000}. In this setting, note that the best $\hat K$ is the one which ensures better classification performances of the corresponding CFA\texttt{+}EFA model compared to models with other $\hat K$ values. To this end,	 the entropy measure has demonstrated to be particularly appropriate for the CFA\texttt{+}EFA model, as reported in a small simulation study (see the supplementary materials). 
{However, some suggestions on the choice of $K$ can be drawn form theoretical/empirical considerations (Section \ref{sec:2}), simulations (Section \ref{sec:4}), and case studies (Section \ref{sec:5}). To treat faking, an optimal choice has been demonstrated to be $K=1$, whereas careless responding seems to require a tailor-made model selection by varying the $K$ values.}

\subsection{Parameter estimation}

Maximum likelihood estimates of $\boldsymbol{\Omega}$ are usually obtained via the EM algorithm \cite{dempster1977maximum}, which entails the maximization of the complete log-likelihood function, i.e. the joint p.d.f. of the observed and latent data. Let $\boldsymbol{u}=\{\boldsymbol{\eta},\boldsymbol{\xi},\boldsymbol{z}\}$ be the set of latent variables, the complete log-likelihood function is:
\begin{align}
	\ell(\boldsymbol{\Omega};\boldsymbol{y},\boldsymbol{u})&=\log\Bigg(\prod_{i=1}^n f\left(\boldsymbol{y}_i\mid\{\boldsymbol{\eta}_i,z_i=1\}\right)^{z_i} f\left(\boldsymbol{\eta}_i\mid \{z_i=1\}\right)^{z_i} f({z}_i=1)^{z_i} \times\\\nonumber
	&\quad\quad\quad\times\prod_{i=1}^n f\left(\boldsymbol{y}_i\mid\{\boldsymbol{\xi}_i,z_i=0\}\right)^{1-z_i} f\left(\boldsymbol{\xi}_i\mid \{z_i=0\}\right)^{1-z_i} f({z}_i=0)^{1-z_i}\Bigg)
\end{align}
where, for the sake of simplicity, parameters within the distributions of the observed and latent variables have been omitted. In this form, the CFA\texttt{+}EFA model requires the estimation of $3p+pK+q+K+C+1+q(q-1)/2$ parameters. In particular, the EM algorithm provides an iterative procedure which generates a sequence of non-decreasing lower bounds for the maximization of $\mathbb{Q}(\boldsymbol{\Omega},\hat{\boldsymbol{\Omega}})=\mathbb{E}_{}\left[\ell\left(\boldsymbol{\Omega};\boldsymbol{y},\boldsymbol{u}\right)\mid \boldsymbol{y}\right]$, i.e. the complete log-likelihood given the observed data \cite{mclachlan2007algorithm}. Thus, a generic iteration of the EM algorithm requires the calculation of the expectation $\mathbf{\hat u} =\mathbb{E}_{}\left[\ell\left(\boldsymbol{\Omega};\boldsymbol{y},\boldsymbol{u}\right)\mid \boldsymbol{y}\right]$, which is in turn plugged into a Newton method to complete the maximization problem for the mixture parameter. Overall, for a given iteration $t$ the EM algorithm proceeds as follows:
\begin{enumerate}
	\item \textbf{E-step}: compute $\mathbf{\hat u} = \mathbb{E}_{}\left[\ell\left(\boldsymbol{\Omega};\boldsymbol{y},\boldsymbol{u}\right)\mid \boldsymbol{y}\right]$ given $\boldsymbol{\Omega}^{(t-1)}$ \vspace{0.15cm}
	\item \textbf{M-step}: $\hat{\boldsymbol{\Omega}}^{(t)}=\arg\underset{\boldsymbol{\Omega}}{\max}\,\mathbb{Q}(\boldsymbol{\Omega},\hat{\boldsymbol{\Omega}}^{(t-1)})$ using $\mathbf{\hat u}$
	\item For $\epsilon>0$, convergence if $\epsilon>\,\mid\!\mathbb{Q}(\hat{\boldsymbol{\Omega}}\mid\hat{\boldsymbol{\Omega}}^t)-\mathbb{Q}(\hat{\boldsymbol{\Omega}}\mid\hat{\boldsymbol{\Omega}}^{(t-1)})\!\mid$
\end{enumerate}
All technical derivations and details of the EM algorithm are reported as online supplementary material to this article.

\section{Simulation studies}\label{sec:4}

\subsection{Simulation study 1}\label{sec:4.1}

The aim of this simulation study was to assess the classification performances of the CFA\texttt{+}EFA model, with a particular focus on understanding the impact of varying proportions $\boldsymbol{\pi}$ within the observed sample. For the sake of comparison, we have also considered the extreme condition $\pi=0.05$, where the aberrant observations constitute the 95\% of the overall sample. 

\paragraph{\textit{Design}}
The design of this simulation study involved four factors: (i) $\pi\in\{0.05,0.40,0.60,0.80,0.90\}$, (ii) $q\in\{1,3\}$, (iii) $K\in\{2,4\}$, (iv) $C\in\{1,2\}$. The factors were systematically varied in the complete factorial design with a total of $5\times2\times2\times2=40$ scenarios. The invariant inputs for the model were the sample size fixed at $n=1000$ and the number of observed variables $p=30$. For each combination, $B=1000$ samples were generated which correspond to $1000\times40=40000$ new data and an equivalent number of parameters.

\paragraph{\textit{Procedure}}
Let $q_h$, $\pi_w$, $k_s$, and $c_m$ be different levels of factors $q$, $\pi$, $K$, and $C$. Then, the data matrices were generated as follows:

\begin{itemize}
	\item[(a)] For $i=1,\dots,n$, $j=1,\dots,p$, $q=1,\dots,q_l$ and $k=1,\dots,k_s$, the true parameters of the CFA model \cite{brown2015cfa}  were obtained as:
	\begin{align*}
		&\lambda_{1j\times q_{l}}\sim \text{U}(0.05,0.99),\quad \boldsymbol{\Phi}_{q_l\times q_l}\sim\text{LKJ}(1,q_{l}),\quad \boldsymbol{\Lambda}_{1p\times q_{l}} = \boldsymbol{\Lambda}_{1p\times q_l}\odot\boldsymbol{\Lambda}^{\text{str}}_{p\times q_l},\quad	\boldsymbol{\mu} = \boldsymbol{0}_{q_l}\\[0.10cm]
		&\boldsymbol{\Theta}_{\boldsymbol{\delta}p\times p}=\boldsymbol{1}_{p\times1}-\text{diag}(\boldsymbol{\Lambda}_{p\times q_l}\boldsymbol{\Phi}_{q_l\times q_l}\boldsymbol{\Lambda}_{p\times q_l}^T),\quad\lambda_{2j\times k}\sim \text{U}(0.05,0.99),\quad \quad 		\boldsymbol{\Psi}_{\boldsymbol{\epsilon}p\times p}=0.85\cdot \mathbf{I}_{p\times p},\\[0.10cm]
		&\beta_1 = \log\left(\frac{\pi_w}{\left(1 - \pi_w\right)}\right), \quad \boldsymbol{\beta}_{c_m+1} \sim \text{U}(-1.5,1.5),\quad x_{i,0+1} = 1,\quad x_{i,1+1} \sim \mathcal{B}ern(0.5),\\[0.10cm]
		& \boldsymbol{\nu}_{k_{s}\times1}\sim \text{U}(0.5,5), \quad x_{i,2+1}\sim \text{U}(-5,5)
	\end{align*}
	\noindent where $\text{U}(0.05,0.99)$ represents a continuous uniform distribution with a closed interval $[0.05,0.99]$, $\text{LKJ}(1,q_l)$ indicates the Lewandowski-Kurowicka-Joe distribution with shape parameter equals to 1 and dimension $q_l\times q_l$, whereas $\boldsymbol{\Lambda}^{\text{str}}_{p\times q_l}$ is a matrix of zeros and ones which determines the constrained structure of $\boldsymbol{\Lambda}_{1p\times q_l}$ (with no cross-loadings). \vspace{0.15cm}
	\item[(b)] For $m = 1,\dots,M$ with $M=50000$, the CFA respondents' latent traits, measurement errors, and continuous vectors of responses were computed as $\boldsymbol{\eta}_{q_l\times m}\sim N_{q_l}(\boldsymbol{0}_q, \boldsymbol{\Phi}_{q_l\times q_l})$, $\boldsymbol{\delta}_{p\times m}\sim N_{p}(\boldsymbol{0}_{p},\boldsymbol{\Theta}_{\boldsymbol{\delta}p\times p})$, and $\boldsymbol{y}_{p\times m} = \boldsymbol{\Lambda}_{1p\times q_l}\boldsymbol{\eta}_{ q_l\times m}+\boldsymbol{\delta}_{p\times m}$. \vspace{0.15cm}
	\item[(c)] For $r = 1,\dots, R$ with $R=50000$, the EFA respondents' latent traits, measurement errors, and continuous vectors of responses were computed as $\boldsymbol{\xi}_{k_l\times r}\sim N_{k_l}(\boldsymbol{0}_{k_l}, \boldsymbol{I}_{k_l\times k_l})$, $\boldsymbol{\epsilon}_{p\times r}\sim N_{j}(\boldsymbol{0}_{p},\boldsymbol{\Psi}_{\boldsymbol{\epsilon}p\times p})$, and $\boldsymbol{y}_{p\times r} = \boldsymbol{\Lambda}_{2p\times k_l}\boldsymbol{\xi}_{ k_l\times l}+\boldsymbol{\epsilon}_{p\times r}$.	\vspace{0.15cm}
	\item[(d)] For $i = 1,\dots,n$ and $d=c_m+1$, the latent indicator variable of the mixture was sampled by: $\boldsymbol\pi_{n\times1}= {\exp(\boldsymbol{X}_{n\times d}\boldsymbol{\beta}_{d\times1})}/{(\exp(\boldsymbol{X}_{n\times d}\boldsymbol{\beta}_{d\times1})+\boldsymbol{1}_{n\times1})}$ and  $\boldsymbol{z}_{n\times1}\sim \mathcal{B}ern(\pi_{n\times 1})$.\vspace{0.15cm}
	\item[(e)] The data matrix is obtained as $\boldsymbol{Y}_{p\times n}=\left[\{\boldsymbol{y}_{p\times m}\mid \boldsymbol{z}_{i\times1}=\boldsymbol{1}_{i\times1}\},\{\boldsymbol{y}_{p\times r}\mid \boldsymbol{z}_{i\times1}=\boldsymbol{0}_{i\times1}\}\right]^T$ (stacked matrix).
\end{itemize}

\paragraph{\textit{Classification measures}}
To evaluate the classification performances of the model, we considered the following {classification metrics (CM) }\cite{luque2019impact, lopez2013insight}: (i) \textit{sensitivity} (SE),  which ranges between 0 (completely inaccurate) and 1 (completely accurate); (ii) \textit{specificity} (SP), which ranges between 0 (completely inaccurate) and 1 (completely accurate); (iii) \textit{balanced accuracy} (BACC), which serves as an indicator of  how different the classification is from random guessing and varies between 0 (i.e., perfect misclassification) and 1 (i.e., perfect classification); (iv)  \textit{Matthews Correlation Coefficient} (MCC), which corresponds to the geometric means of eight ratios derived from the combinations of all the components of a confusion matrix and is a good metric for representing the global model quality \cite{zhu2020performance}, ranging between -1 (completely inaccurate) and 1 (completely accurate), with 0 equal to the random guessing. Considering the number of correctly detected biased respondents (i.e., \textit{true positives}, TP), incorrectly flagged unbiased respondents (i.e., \textit{false positives}, FP), correctly detected unbiased respondents (i.e., \textit{true negatives}, TN), incorrectly not flagged  biased respondents (i.e., \textit{false negatives}, FN), these metrics are formally defined as follows:
\begin{align*}
	&\text{SE}= \frac{\text{TP}}{\text{TP}+\text{FN}},\quad\text{SP}=\frac{\text{TN}}{\text{TN}+\text{FP}},\quad\text{BACC} = \frac{\left(\text{SE}+\text{SP}\right)}{2},\\[0.10cm]
	&\text{MCC} = \frac{\text{TP}\times\text{TN}-\text{FP}\times\text{FN}}{\sqrt{(\text{TP}+\text{FP})(\text{TP}+\text{FN})(\text{TN}+\text{FP})(\text{TN}+\text{FN})}}
\end{align*}

\paragraph{\textit{Results and discussion}}

Table \ref{tab:1} reports the classification results averaged along with standard deviations. Overall, according to classification properties of mixture models \cite{mcnicholas2016mixture}, the CFA\texttt{+}EFA model reports satisfactory results. With regard to $C$, the indices generally show larger values for $C=2$, which suggests that including additional quantitative information in terms of covariates lead to an improvement of the classification performances. Similarly, with regard to ${\pi}$, all the indices show larger values when the proportion of aberrant respondents is not so large (i.e., ${\pi}\geq0.60$). In this case, as expected, BACC and MCC show a slightly decreasing trend when $C=1$, and an increasing trend when $C=2$. On the other hand, the SP measure tends to increase independently of covariates. The slight decrease in the conditions $C=1$ can be motivated by a slightly diminished identification of the aberrant respondents if compared to the $C=2$ conditions.  With respect to $K$, the condition $K=4$ reports higher values if compared to $K=2$, suggesting improved discrimination for those models with a higher number of EFA latent factors. Interestingly, for $q=3$, the condition $K=2$ presents slightly higher values if compared to $K=4$, while the opposite is true when $q=1$. This result is not surprising, as these conditions involve more substantial separation between the CFA and EFA parametric spaces. {Finally, looking at conditions $q=1$ with respect to $K=\{2,4\}$, the results indicate that, when the true number of EFA factors $K$ is not in the neighbourhood of $q$, the discrimination between the CFA and EFA observations improves, which was expected since more dissimilarity between the mixture components contributes to better classifications.}

\addtolength{\tabcolsep}{4pt}   
\begin{table}[!t]
	\caption{Monte Carlo study 1: average balanced accuracy (BACC), Matthew's correlation coefficient (MCC), sensitivity (SE), and specificity (SP) and  the corresponding standard deviations across the replications.\label{tab:1}}
	
	\resizebox{!}{3.53cm}{
		\begin{tabular}{ccccccccccc}
			\toprule
			\multirow{2}{*}{$\pi$}&\multirow{2}{*}{$K$}&\multirow{2}{*}{$q$}&\multicolumn{4}{c}{$C=1$}&\multicolumn{4}{c}{$C=2$}\\\cmidrule(lr){4-7}\cmidrule(lr){8-11}
			& &&  $\text{BACC}$  & $\text{MCC}$  & $\text{SE}$ & $\text{SP}$&   $\text{BACC}$ & $\text{MCC}$  & $\text{SE}$ & $\text{SP}$\\
			\midrule
			$0.05$ & $2$ & $1$ & $0.822\,(0.026)$ & $0.599\,(0.042)$ & $0.938\,(0.008)$ & $0.705\,(0.053)$ & $0.941\,(0.01$) & $0.883\,(0.017)$ & $0.973\,(0.006)$ & $0.908\,(0.021)$\\
			&&$3$ &$0.858\,(0.038)$ & $0.666\,(0.065)$ & $0.977\,(0.005)$ & $0.738\,(0.075)$ & $0.805\,(0.052)$ & $0.536\,(0.085)$ & $0.973\,(0.005)$ & $0.637\,(0.101)$\\[0.10cm]
			& $4$ & $1$ & $0.8\,(0.052)$ & $0.581\,(0.081)$ & $0.984\,(0.004)$ & $0.616\,(0.103)$ & $0.953\,(0.009)$ & $0.91\,(0.014)$ & $0.98\,(0.005)$ & $0.927\,(0.017)$\\
			&&$3$ &$0.732\,(0.059)$ & $0.44\,(0.104)$ & $0.982\,(0.004)$ & $0.483\,(0.116)$ & $0.918\,(0.016)$ & $0.828\,(0.025)$ & $0.973\,(0.006)$ & $0.863\,(0.031)$\\[0.20cm]
			$0.40$ & $2$ & $1$ &$0.899\,(0.009)$ & $0.8\,(0.019)$ & $0.923\,(0.014)$ & $0.876\,(0.017)$ & $0.901\,(0.01$) & $0.803\,(0.018)$ & $0.929\,(0.011)$ & $0.872\,(0.019)$\\
			&&$3$ &$0.941\,(0.008)$ & $0.879\,(0.016)$ & $0.944\,(0.012)$ & $0.938\,(0.011)$ & $0.96\,(0.005)$ & $0.92\,(0.011)$ & $0.961\,(0.009)$ & $0.959\,(0.01$)\\[0.10cm]
			& $4$ & $1$ & $0.946\,(0.007)$ & $0.891\,(0.014)$ & $0.958\,(0.01$) & $0.933\,(0.012)$ & $0.967\,(0.005)$ & $0.935\,(0.01$) & $0.974\,(0.007)$ & $0.96\,(0.01$)\\
			&&$3$ &$0.903\,(0.012)$ & $0.805\,(0.023)$ & $0.941\,(0.011)$ & $0.865\,(0.022)$ & $0.882\,(0.113)$ & $0.735\,(0.241)$ & $0.821\,(0.167)$ & $0.944\,(0.103)$\\[0.20cm]
			$0.60$ & $2$ & $1$ &$0.951\,(0.008)$ & $0.884\,(0.017)$ & $0.942\,(0.016)$ & $0.959\,(0.008)$ & $0.909\,(0.009)$ & $0.82\,(0.018)$ & $0.927\,(0.013)$ & $0.891\,(0.016)$\\
			&&$3$ &$0.908\,(0.009)$ & $0.808\,(0.019)$ & $0.91\,(0.016)$ & $0.906\,(0.012)$ & $0.962\,(0.006)$ & $0.924\,(0.011)$ & $0.966\,(0.008)$ & $0.958\,(0.01$)\\[0.10cm]
			& $4$ & $1$ & $0.946\,(0.009)$ & $0.869\,(0.018)$ & $0.937\,(0.018)$ & $0.956\,(0.008)$ & $0.984\,(0.004)$ & $0.968\,(0.007)$ & $0.985\,(0.005)$ & $0.983\,(0.006)$\\
			&&$3$ &$0.951\,(0.008)$ & $0.89\,(0.015)$ & $0.944\,(0.015)$ & $0.957\,(0.008)$ & $0.963\,(0.006)$ & $0.925\,(0.012)$ & $0.964\,(0.009)$ & $0.961\,(0.009)$\\[0.20cm]
			$0.80$ & $2$ & $1$ & $0.92\,(0.015)$ & $0.786\,(0.027)$ & $0.888\,(0.03$) & $0.952\,(0.007)$ & $0.965\,(0.006)$ & $0.929\,(0.011)$ & $0.962\,(0.01$) & $0.969\,(0.008)$\\
			&&$3$ &$0.912\,(0.01$) & $0.808\,(0.02$) & $0.901\,(0.02$) & $0.924\,(0.01$) & $0.946\,(0.008)$ & $0.885\,(0.015)$ & $0.93\,(0.015)$ & $0.962\,(0.008)$\\[0.10cm]
			& $4$ & $1$ & $0.944\,(0.011)$ & $0.856\,(0.021)$ & $0.928\,(0.021)$ & $0.959\,(0.007)$ & $0.956\,(0.008)$ & $0.898\,(0.016)$ & $0.946\,(0.015)$ & $0.966\,(0.007)$\\
			&&$3$ &$0.948\,(0.009)$ & $0.881\,(0.017)$ & $0.937\,(0.019)$ & $0.959\,(0.007)$ & $0.959\,(0.007)$ & $0.915\,(0.014)$ & $0.955\,(0.013)$ & $0.963\,(0.008)$\\[0.20cm]
			$0.90$ & $2$ & $1$ &$0.885\,(0.024)$ & $0.685\,(0.04$) & $0.82\,(0.048)$ & $0.949\,(0.007)$ & $0.952\,(0.007)$ & $0.903\,(0.014)$ & $0.935\,(0.014)$ & $0.969\,(0.007)$\\
			&&$3$ &$0.915\,(0.031)$ & $0.736\,(0.051)$ & $0.851\,(0.062)$ & $0.978\,(0.004)$ & $0.924\,(0.011)$ & $0.833\,(0.02$) & $0.901\,(0.021)$ & $0.946\,(0.008)$\\[0.10cm]
			& $4$ & $1$ & $0.905\,(0.03$) & $0.764\,(0.048)$ & $0.826\,(0.06$) & $0.983\,(0.004)$ &  $0.990\,(0.003)$ & $0.979\,(0.007)$ & $0.989\,(0.006)$ & $0.992\,(0.003)$\\
			&&$3$ &$0.932\,(0.026)$ & $0.821\,(0.041)$ & $0.881\,(0.051)$ & $0.983\,(0.004)$ & $0.965\,(0.014)$ & $0.903\,(0.025)$ & $0.942\,(0.027)$ & $0.989\,(0.003)$\\
			\bottomrule
	\end{tabular}}
\end{table}
\addtolength{\tabcolsep}{-4pt}

\subsection{Simulation study 2}\label{sec:4.2}

The aim of this simulation study was twofold: to externally validate the CFA\texttt{+}EFA model for detecting faked observations and to assess the quality of the predicted unbiased data. To simulate a realistic perturbation of the data matrix, we used the SGR model which is an empirically validated psychometric model of faking \cite{lombardi2012sensitivity}. To this end, we followed a two-step procedure. Initially, we  generated rating data based on a standard CFA model. Subsequently, we selected a subset of observations to be subject to faking, mimicking extreme and slight styles of faking using the SGR model. 

\paragraph{\textit{Data perturbation model}} 
Discrete rating data $Y_{ij}\in\{1,\dots m,\dots,M\}$ can be generated by sampling the fake response pattern $\boldsymbol{f}_i$ from the so-called \textit{conditional replacement distribution}, given the uncorrupted observations $\boldsymbol{y}_i$ \cite{lombardi2015sgr}, which takes the following form:
\begin{equation*}
	P(f_{ij}=m'\mid y_{ij}=m,\gamma,\delta) = 
	\begin{cases}
		1& m=m'=M\\
		\kappa\text{DG}(m';m+1,M,\gamma,\delta), & 1\leq m<m'\leq M\\
		1-\kappa, & 1\leq m'=m< M\\
		0, & 1\leq m'< m\leq M\\
	\end{cases}
\end{equation*}
where $m'$ is the replaced (faked) response, $\kappa$ denotes the overall probability of faking good, $\text{DG}(m';m+1,M,\gamma,\delta)$ is the discretized beta distribution with bounds $a = m+1$  and $b = M$, and $\gamma$ and $\delta$ are strictly positive shape parameters of the faking subsample distribution. Note that the model can mimic slight faking good behaviours by setting $\gamma = 1.5$ and $\delta = 4$, and extreme faking good behaviours by setting $\gamma = 4$ and $\delta = 1.5$.

\paragraph{\textit{Design}}
The study involved four factors: (i) $n\in\{250,1000\}$, (ii) $p\in\{16,40\}$, (iii) $\pi\in\{0.6,0.8\}$, (iv) $\{\gamma,\delta\}=\left\{\{1.5,4\},\{4,1.5\}\right\}$.
They were systematically varied in the complete factorial design with a total of $2\times2\times2\times2=16$ scenarios. For each combination, $B=500$ samples were generated which correspond to $500\times16=8000$ new data and an equivalent number of parameters. For the sake of simplicity, some other conditions were not varied: (i) the CFA latent factors were fixed to $q=4$, which reproduces a quite common latent test dimensionality; (ii) the EFA latent factors were set to $K=1$, which represents a reasonable latent structure for those respondents adopting faking behaviours. 

\paragraph{\textit{Procedure}}
Let $i_v$, $j_h$, $\pi_w$, $\{\gamma_o,\delta_o\}$ be different levels of factors $n$, $p$, $\pi$, $\{\gamma,\delta\}$. Then, the perturbed data matrix were generated as follows:

\begin{itemize}
	\item[(a)] For $j=1,\dots,j_h$, the true parameters of the CFA model \cite{brown2015cfa}  were obtained by letting:
	\begin{align*}
		&\lambda_{j\times q}\sim \text{U}(0.05,0.99),\quad \boldsymbol{\Phi}\sim\text{LKJ}(1,q),\quad \boldsymbol{\Lambda}_{j_h\times q} = \boldsymbol{\Lambda}_{j_h\times q}\odot\boldsymbol{\Lambda}^{\text{str}}_{j_h\times q},\\[0.10cm]
		&\boldsymbol{\Theta}_{\boldsymbol{\delta}j_h\times j_h}=\boldsymbol{1}_{j_h\times1}-\text{diag}(\boldsymbol{\Lambda}_{j_h\times q}\boldsymbol{\Phi}\boldsymbol{\Lambda}_{j_h\times q}^T)\quad\boldsymbol{\mu} = \boldsymbol{0}_q,\quad
		\boldsymbol{\beta} = \log\left(\frac{\pi_w}{\left(1 - \pi_w\right)}\right)
	\end{align*}
	\noindent where $\text{U}(0.05,0.99)$ represents a continuous uniform distribution with a closed interval $[0.05,0.99]$, $\text{LKJ}(1,q)$ indicates the Lewandowski-Kurowicka-Joe distribution with shape parameter equals to 1 and dimension $q\times q$, and $\boldsymbol{\Lambda}^{\text{str}}_{j_h\times q}$ is a matrix of zeros and ones which determines the constrained structure of the $\boldsymbol{\Lambda}_{j_h\times q}$, by fixing to zeros some factor loadings. \vspace{0.15cm}
	\item[(b)] Respondents' latent traits, measurement errors, and continuous vectors of responses were computed as $\boldsymbol{\eta}_{q\times i_v}\sim N_q(\boldsymbol{0}_q, \boldsymbol{\Phi})$, $\boldsymbol{\delta}_{j_h\times i_v}\sim N_{j_h}(\boldsymbol{0}_{j_h},\boldsymbol{\Theta}_{\boldsymbol{\delta}j_h\times j_h})$, and $\boldsymbol{y}_{j_h\times i_v}^* = \boldsymbol{\Lambda}_{j_h\times q}\boldsymbol{\eta}_{ q\times i_v}+\boldsymbol{\delta}_{j_h\times i_v}$. \vspace{0.15cm}
	\item[(c)] For $i=1,\dots,i_v$ and $j=1,\dots,j_h$, the continuous data $\boldsymbol{y}_{j\times i}^*$ were discretized into $M=11$ response categories:
	\begin{align*}
		y_{j\times i}&=m\Longleftrightarrow\tau_{m-1}<y^*_{ji}<\tau_m, \,\, m=1,2,\dots,M\\
		\tau_0&=-\infty, \, \tau_1<\tau_2<\dots<\tau_{M-1},\,\, \tau_{M}=+\infty
	\end{align*}
	\item[(d)]  Let $n_{f}=n_v\cdot\left(1-\pi_w\right)$ respectively denotes the observations in the faking subsample, the matrix of responses to be perturbed was sampled by $\boldsymbol{z}_{i_v\times1}\sim \mathcal{B}ern(\pi_w)$ and $\boldsymbol{D}_{j_h\times n_f}=\{\boldsymbol{Y}_{j_h\times i_v}\mid \boldsymbol{z}_{i_v\times1}=\boldsymbol{0}_{i_v\times1}\}$. \vspace{0.15cm}
	\item[(e)] The SGR conditional replacement distribution with faking good probability $\kappa=1$ is:
	\begin{equation*}
		P(f_{ji}=m'\mid d_{ji}=m,\gamma,\delta) = 
		\begin{cases}
			1, & m=m'=M\\
			\text{DG}(m';m+1,M,\gamma_o,\delta_o), & 1\leq m<m'\leq M\\
			0, & 1\leq m'\leq m\leq M
		\end{cases}
	\end{equation*}
	\item[(f)] Let $n_{f}=n_v\cdot(1-\pi_w)$ and $n_{d}=n_v\cdot \pi_w$ denoting the observations in the faking subsample and in the unbiased sample. The complete data matrix was obtained by $\boldsymbol{Y}_{j_h\times i_v} = 	\left[\boldsymbol{F}_{j_h\times n_f},		\boldsymbol{D}_{j_h\times n_d}\right]^T$ (stacked matrix).
\end{itemize}

\paragraph{\textit{Measures}}
In this study, all the classification measures described in Simulation Study 1 were used along with the bias and the root mean square error (RMSE) of estimates for assessing the ability of the algorithm to recover the true model parameters.

\paragraph{\textit{Results and discussion}}

Table \ref{tab:2} reports the classification results averaged along with standard deviations. As expected, all the indices show good classification properties in cases of \texttt{extreme} faking. In general, all the classification metrics - except SE - show a better performance with a larger sample size ($n=1000$) and a higher proportion of CFA-related observations ($\pi=0.80$). The SE results may corroborate that these mixture models may suffer of lower SE when reconstructing the component memberships in presence of a lower number of observations in one of the two mixture components \cite{cintron2023cautionary}. With regard to ${p}$, the condition $p=16$ tends to present better performances in the case of \texttt{slight} faking, but the opposite is true in the case of \texttt{extreme} faking with a higher number of items ($p=30$), which results in an improvement of performances across all the indices. Overall, the results indicate that the CFA\texttt{+}EFA model is a satisfactory classifier for detecting observations affected by an extreme degree of faking, { which is highly detrimental to many statistical tools commonly used in data analysis} \cite{bressan2018effect,lee2019investigating,pastore2014impact}. 

\begin{table}[!t]
	\caption{Monte Carlo study 2: average of balanced accuracy (BACC), Matthew's correlation coefficient (MCC), sensitivity (SE), and specificity (SP)  and  the corresponding standard deviations across the replications  in parentheses.\label{tab:2}}
	\resizebox{!}{1.79cm}{
		\begin{tabular}{ccccccccccc}
			\toprule
			\multirow{2}{*}{Faking}& \multirow{2}{*}{$\pi$}& \multirow{2}{*}{$p$}& \multicolumn{4}{c}{$n=250$} & \multicolumn{4}{c}{$n=1000$}\\\cmidrule(lr){4-7}\cmidrule(lr){8-11}
			&&&BACC &  MCC & SE & SP & BACC & MCC & SE & SP \\
			\midrule
			\texttt{slight}&$0.60$&$16$&$0.682\,(0.097)$ & $0.308\,(0.152)$ & $0.647\,(0.181)$ & $0.717\,(0.079)$ & $0.745\,(0.090$) & $0.320\,(0.145)$ & $0.829\,(0.156)$ & $0.662\,(0.039)$\\
			&&$30$&$0.690\,(0.080$) & $0.271\,(0.108)$ & $0.720\,(0.147)$ & $0.659\,(0.030$) & $0.769\,(0.052)$ & $0.329\,(0.107)$ & $0.879\,(0.093)$ & $0.658\,(0.026)$\\[0.10cm]
			&$0.80$&$16$&$0.824\,(0.130$) & $0.470\,(0.218)$ & $0.783\,(0.235)$ & $0.864\,(0.039)$ & $0.854\,(0.100)$& $0.641\,(0.135)$ & $0.794\,(0.195)$ & $0.914\,(0.022)$\\
			&&$30$&$0.731\,(0.151)$ & $0.209\,(0.163)$ & $0.643\,(0.294)$ & $0.820\,(0.026)$ & $0.716\,(0.199)$ & $0.332\,(0.369)$ & $0.592\,(0.320$) & $0.841\,(0.097)$\\[0.20cm]
			\texttt{extreme}&$0.60$&$16$&$0.965\,(0.044)$ & $0.935\,(0.093)$ & $0.934\,(0.088)$ & $0.997\,(0.006)$ & $0.980\,(0.033)$ & $0.961\,(0.067)$ & $0.973\,(0.046)$ & $0.987\,(0.025)$\\
			&&$30$&$0.991\,(0.011)$ & $0.984\,(0.020$) & $0.986\,(0.022)$ & $0.996\,(0.007)$ & $0.997\,(0.002)$ & $0.995\,(0.004)$ & $0.996\,(0.004)$ & $0.999\,(0.002)$\\[0.10cm]
			&$0.80$&$16$&$0.981\,(0.062)$ & $0.964\,(0.115)$ & $0.969\,(0.104)$ & $0.993\,(0.022)$ & $0.993\,(0.010$) & $0.989\,(0.015)$ & $0.987\,(0.020$) & $0.999\,(0.001)$\\
			&&$30$&$0.994\,(0.042)$ & $0.985\,(0.078)$ & $0.992\,(0.069)$ & $0.996\,(0.016)$ & $0.998\,(0.006)$ & $0.996\,(0.018)$ & $0.996\,(0.008)$&$1.000\,(0.005)$\\
			\bottomrule
	\end{tabular}}\textcolor{white}{(Section \ref{sec:2})}
	\centering
\end{table}

\noindent Table \ref{tab:3} shows the averaged bias and RMSE values along with standard deviations for both slight and extreme faking conditions. With respect to $\hat{\boldsymbol{\Lambda}}$, the condition $p=16$ shows negligible bias whereas conditions with a higher number of observed indicators ($p=40$) exhibit less accurate recovery. With regard to $\hat{\boldsymbol{\Theta}}_{\boldsymbol{\delta}}$, all the conditions indicate that the recovery bias is almost insignificant and no clear differences emerged among the faking scenarios. Instead, the recovery of $\hat{\boldsymbol{\Phi}}$ is overall inaccurate. With respect to $\hat{\boldsymbol{\mu}}$, as expected, the estimates display substantial bias in the \texttt{extreme} conditions but are accurate in the \texttt{slight} conditions. Finally, the recovery of the mixture parameter $\hat{\pi}$ is generally satisfactory, especially in the case of \texttt{extreme} faking. Overall, these results suggest that estimating the CFA\texttt{+}EFA model in slight faking conditions might be challenging, as in the case of mild alteration of the observed covariance matrix. In this case, instead of being captured by the EFA submodel, the aberrant subsample would simply affect the factorial structure of the CFA submodel and, possibly, would increase its residual variances.

\begin{table}[!t]
	\caption{Monte Carlo study 2: average bias  and average root mean squared error (RMSE) between the true parameters of the original CFA model and  the estimated parameters of the CFA component  in the CFA\texttt{+}EFA model along with the standard deviations across the replications  in parentheses.\label{tab:3}}
	\resizebox{!}{2.37cm}{
		\begin{tabular}{cccccccccccccc}
			\toprule
			\multirow{2}{*}{Faking}&\multirow{2}{*}{$\pi$}&\multirow{2}{*}{$p$}&\multirow{2}{*}{$n$}&\multicolumn{2}{c}{$\hat{\boldsymbol{\Lambda}}$}&\multicolumn{2}{c}{$\hat{\boldsymbol{\Theta}}_{\boldsymbol{\delta}}$}&\multicolumn{2}{c}{$\hat{\boldsymbol{\Phi}}$}&\multicolumn{2}{c}{$\hat{\boldsymbol{\mu}}$}&\multicolumn{2}{c}{$\hat{\pi}$}\\\cmidrule(lr){5-6} \cmidrule(lr){7-8} \cmidrule(lr){9-10} \cmidrule(lr){11-12} \cmidrule(lr){13-14}
			&&&& $\text{Bias}$ & $\text{RMSE}$ & $\text{Bias}$ & $\text{RMSE}$ & $\text{Bias}$ & $\text{RMSE}$ & $\text{Bias}$  & $\text{RMSE}$ & $\text{Bias}$ & $\text{RMSE}$\\
			\midrule
			\texttt{slight}& $0.60$ &$16$ & $250$  & $0.025 \,(0.016)$& $0.076 \,(0.039)$ & $-0.007 \,(0.002)$  & $0.057 \,(0.031)$ & $0.252 \,(0.087)$  & $0.523 \,(0.160)$ & $-0.123 \,(0.092)$ & $0.126 \,(0.088)$ & $0.038 \,(0.185)$ & $0.163 \,(0.095)$\\
			&&&$1000$  & $0.062 \,(0.007)$ & $0.128 \,(0.015)$ & $-0.017 \,(0.001)$  & $0.090 \,(0.005)$  & $0.265 \,(0.038)$  & $0.535 \,(0.076)$ & $-0.075 \,(0.047)$ & $0.076 \,(0.046)$ & $0.242 \,(0.056)$ & $0.242 \,(0.056)$\\[0.10cm]
			&&$40$&$250$  & $-0.381 \,(0.139)$ & $0.871 \,(0.256)$ & $0.000 \,(0.001)$  & $0.040 \,(0.008)$ & $-0.257 \,(0.093)$  & $0.527 \,(0.184)$  & $-0.015 \,(0.045)$ & $0.039 \,(0.027)$ & $0.229 \,(0.092)$ & $0.234 \,(0.080)$\\
			&&&$1000$  & $-0.230 \,(0.002)$ & $0.593 \,(0.004)$ & $0.001 \,(0.000)$  & $0.033 \,(0.001)$ & $-0.193 \,(0.020)$ & $0.394 \,(0.040)$  & $-0.072 \,(0.032)$ & $0.072 \,(0.032)$& $0.284 \,(0.047)$ & $0.284 \,(0.047)$\\[0.15cm]
			&$0.80$&$16$&$250$  & $0.041 \,(0.010)$ & $0.100 \,(0.017)$ & $-0.013 \,(0.001)$  & $0.089 \,(0.009)$ & $0.167 \,(0.045)$ & $0.477 \,(0.065)$ & $-0.083 \,(0.057)$ & $0.084 \,(0.054)$& $0.094 \,(0.048)$ & $0.096 \,(0.044)$\\
			&&&$1000$  & $0.013 \,(0.005)$ & $0.038 \,(0.010)$ & $-0.010 \,(0.001)$  & $0.112 \,(0.010)$ & $0.203 \,(0.026)$ & $0.453 \,(0.051)$ & $-0.143 \,(0.036)$ & $0.143 \,(0.035)$ & $0.002 \,(0.088)$ & $0.080 \,(0.038)$\\[0.10cm]
			&&$40$&$250$  & $-0.246 \,(0.004)$ & $0.609 \,(0.006)$ & $0.002 \,(0.000)$  & $0.038 \,(0.002)$ & $-0.122 \,(0.040)$ & $0.280 \,(0.072)$ & $-0.024 \,(0.034)$ & $0.025 \,(0.033)$ & $0.159 \,(0.045)$ & $0.161 \,(0.040)$\\
			&&&$1000$  & $-0.133 \,(0.067)$ & $0.362 \,(0.139)$ & $-0.001 \,(0.001)$ &  $0.030 \,(0.010)$ & $-0.136 \,(0.189)$ & $0.650 \,(0.124)$  & $-0.002 \,(0.062)$& $0.045 \,(0.043)$ & $-0.031 \,(0.150)$ & $0.104 \,(0.112)$ \\[0.22cm]
			\texttt{extreme}& $0.60$&$16$& $250$  & $0.026 \,(0.050)$  & $0.123 \,(0.094)$ & $-0.016 \,(0.002)$ & $0.104 \,(0.005)$ & $0.076 \,(0.032)$  & $0.311 \,(0.047)$  & $-0.795 \,(0.205)$ & $0.805 \,(0.163)$& $-0.035 \,(0.059)$ & $0.036 \,(0.058)$\\
			&&&$1000$   & $0.011 \,(0.018)$  & $0.182 \,(0.073)$ &$-0.005 \,(0.025)$  & $0.118 \,(0.013)$ & $0.377 \,(0.061)$  & $0.971 \,(0.078)$ & $-1.158 \,(0.116)$ & $1.161 \,(0.089)$ & $-0.007 \,(0.020)$ & $0.011 \,(0.019)$\\[0.10cm]
			&&$40$&$250$   & $-0.125 \,(0.108)$ & $0.339 \,(0.242)$ & $-0.003 \,(0.002)$ & $0.043 \,(0.011)$  & $-0.182 \,(0.182)$& $0.582 \,(0.156)$ & $-1.085 \,(0.407)$ & $1.085 \,(0.407)$ & $-0.004 \,(0.011)$ & $0.007 \,(0.010)$\\
			&&&$1000$   & $-0.169 \,(0.120)$ & $0.471 \,(0.256)$ & $-0.002 \,(0.002)$ & $0.045 \,(0.011)$ & $-0.211 \,(0.245)$ & $0.694 \,(0.184)$ & $-1.016 \,(0.032)$ & $1.016 \,(0.032)$& $-0.001 \,(0.002)$ & $0.002 \,(0.002)$ \\[0.15cm]
			&$0.80$&$16$&$250$   & $0.017 \,(0.010)$ & $0.086 \,(0.020)$ & $-0.021 \,(0.001)$ & $0.106 \,(0.004)$ & $0.220 \,(0.066)$& $0.599 \,(0.106)$ & $-0.580 \,(0.094)$  & $0.582 \,(0.086)$ & $-0.002 \,(0.016)$ & $0.008 \,(0.014)$\\
			&&&$1000$  & $0.005 \,(0.006)$ & $0.069 \,(0.012)$ & $-0.017 \,(0.001)$ & $0.092 \,(0.004)$ & $0.124 \,(0.027)$ & $0.362 \,(0.038)$ & $-0.639 \,(0.054)$ & $0.639 \,(0.054)$ & $-0.005 \,(0.008)$ & $0.005 \,(0.008)$\\[0.10cm]
			&&$40$&$250$   & $-0.258 \,(0.004)$ & $0.633 \,(0.008)$ & $-0.001 \,(0.000)$ & $0.030 \,(0.001)$ & $-0.234 \,(0.047)$ & $0.485 \,(0.066)$  & $-0.485 \,(0.047)$& $0.485 \,(0.046)$& $0.003 \,(0.014)$  & $0.003 \,(0.014)$\\
			&&&$1000$   & $-0.097 \,(0.061)$ & $0.284 \,(0.123)$ & $-0.005 \,(0.001)$ & $0.040 \,(0.005)$ & $-0.227 \,(0.054)$ & $0.634 \,(0.096)$ & $-0.048 \,(0.193)$& $0.128 \,(0.151)$& $-0.001 \,(0.004)$  & $0.001 \,(0.004)$ \\
			\bottomrule
	\end{tabular}}	
	\centering
\end{table}

\section{Case studies}\label{sec:5}

In this section, two applications to real datasets will be presented and discussed. The first dataset (first case study) was collected under a controlled scenario by means of which faking behaviour was experimentally manipulated, while the second dataset (second case study) have been found to be affected by careless responding (see \cite{arias2020little}). Both cases serve as a way to assess the empirical validity of the CFA\texttt{+}EFA model, in particular whether it can be used as a general method for detecting respondents who exhibit aberrant rating behaviours. 

\subsection{Case study 1: Faking behaviour} \label{sec:5.1}

\paragraph{\textit{Data and measures}}
{Data were originally collected by \cite{krammer2020applicant} and are available at \url{https://osf.io/wdv25/}. The original dataset consists of $n=763$  teacher-education students (69.33\% females, ages ranged from 18 to 47, with mean of 21.11 and standard deviation of 3.60) who took part to a personality survey  based on the administration of the  Big Five Inventory (BFI), which is a five-factor personality questionnaire composed by $p=42$ items scored on a five-point scale ranging from 1 (``strongly disagree") to 5 (``strongly agree") \cite{lang2001testgute}. The BFI was administered in a honest vs. an inducted faking conditions. In the honest condition, the respondents were asked to compile the BFI according to the standard instructions (no faking), while in the faking condition (after ten months) the same participants were instructed to adopt a faking behaviour when rating the items.  By randomly extracting different participants from the two conditions, two subsamples consisting of both unbiased (no fakers) and biased (fakers) respondents were obtained and combined. This process resulted in the final rating matrix  $\boldsymbol{Y}_{763\times 42}$. Since faking behaviour has been found to vary as a function of age differences \cite{thumin1993faking}, the variable \texttt{age} has been used as predictor of $\boldsymbol{\pi}$, which controls the proportion of non-fakers in this case.}

\begin{figure}[!t]
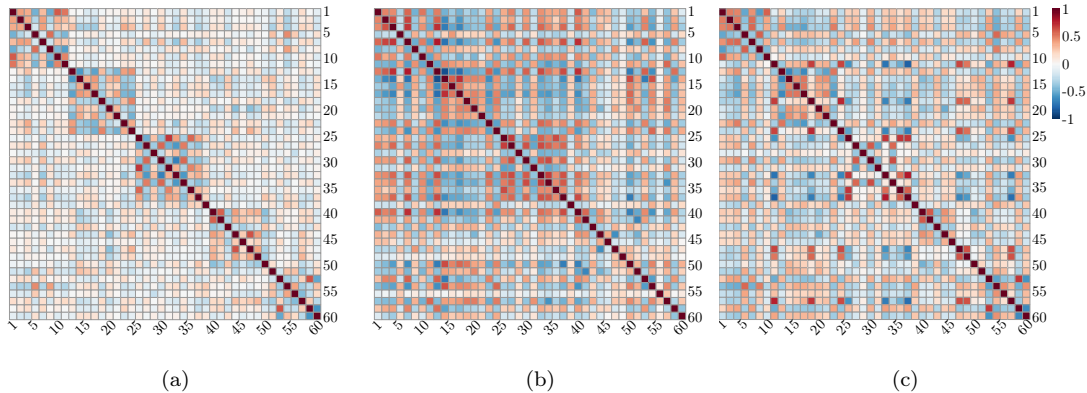

	\hspace{-0.0cm}
	\subfloat[][]{
		\resizebox{4.95cm}{!}{\input{fig_nonfaking_heat.tex}}\label{hon_heat}
	}\hspace{-0.45cm}
	\subfloat[][]{
		\resizebox{4.95cm}{!}{\input{fig_faking_heat.tex}}\label{fak_heat}
	}\hspace{-0.45cm}
	\subfloat[][]{
		\resizebox{4.95cm}{!}{\input{fig_tot_heat.tex}}\label{fak_tot}
	}
	\caption{Heatmaps of the inter-item correlation matrices. Panel (a) represents the  honest subsample's correlation matrix; Panel (b) the faking subsample's correlation matrix; Panel (c) shows the overall sample's correlation matrix.}
	\label{fig_heat_fak}
\end{figure}

\begin{figure}[!t]
	\centering
	\resizebox{7.6cm}{!}{\input{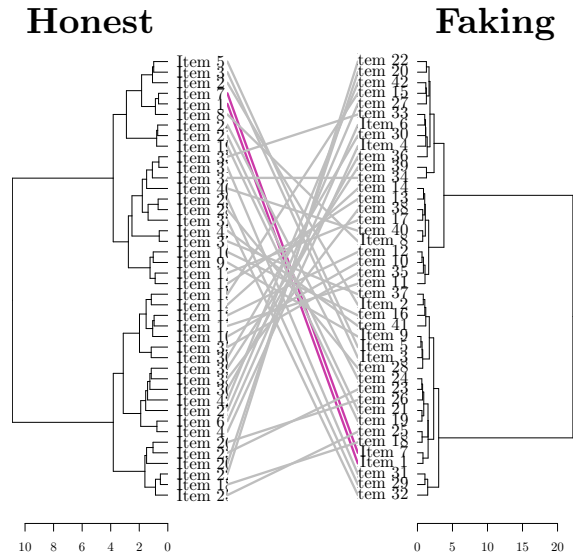}}%
	\caption{Case study 1: Tanglegram  contrasting the dendrograms of  the honest (left) and faking (right) subsamples. The coloured lines indicate common sub-trees between the dendrograms.}
	\label{fig_tangl_fak}
\end{figure}

\paragraph{\textit{Exploratory analyses}}
To justify a CFA\texttt{+}EFA analysis, we assessed the inter-item correlations in the honest and faking conditions. Figure \ref{fig_heat_fak} shows three heatmaps: Figure \ref{hon_heat} represents the  honest subsample's inter-items correlation matrix, Figure \ref{fak_heat} the correlation matrix for the faking subsample, and Figure \ref{fak_tot} shows the overall correlation matrix. By visually inspecting the plots, one can notice differences emerging among the correlation patterns of the subsamples. In particular, the honest subsample seems to overall reproduce the population's correlation matrix, which consist of the five-factor latent structure (i.e., correlation blocks along the main diagonal). By contrast, the subsample of fakers presents a non-random patter of biased inter-item correlations (i.e., a non negligible correlation patterns outside the blocks over the main diagonal), which substantially deteriorate the BFI latent structure.
\noindent To investigate the discrepancies in correlation patterns between honest and faking subsamples, a \textit{tanglegram} was additionally used (see Figure \ref{fig_tangl_fak}). It contrasts the {dendrograms} of honest (on the left) and faking (on the right) subsamples. Dendrograms were constructed by means of a hierarchical clustering based on the Ward's distance. The tanglegram highlights a substantial lack of correspondence between the two dendrograms. To quantify the misalignment between the two dendrograms, we computed the \textit{entanglement coefficient}, which ranges from 0 (full alignment) to 1 (total mismatch). In this case, the entanglement coefficient was found to be $0.76$, confirming a substantial mismatch between the clustered items in the honest and faking subsamples. 

\noindent To verify the impact of faking behaviour on a standard CFA model, we defined, estimated, and evaluated a CFA model based on the BFI latent structure (see \cite{lang2001testgute}) on the overall sample. The results were evaluated by two common fit indices for CFA applications: the comparative fit index (CFI; \cite{bentler1990comparative}), and  the root mean squared error of approximation (RMSEA; \cite{steiger1980statistically}). The analyses were conducted in the \texttt{R} framework \cite{rsoft}  using the \texttt{lavaan} package \cite{rosseel2012lavaan}. The estimated CFA yielded $\text{CFI}=0.543$ (accepted threshold: $\geq 0.90$) and $\text{RMSEA}=0.117$ (accepted threshold: $\leq 0.065$), which suggest that faking has a non-negligible and detrimental impact on the observed correlation matrix as well as on the BFI latent structure.

\paragraph{\textit{CFA\texttt{+}EFA analysis}}

The CFA\texttt{+}EFA model has been specified such that the CFA submodel reproduces the BFI latent structure. In this way, the EFA submodel is left to capture those observations which do not adhere to the population's BFI representation. Note that the constraint $\boldsymbol{\mu}=\boldsymbol{0}_q$ has been used to improve identifiability of the models. Table \ref{tab:4} reports the results of the model selection procedure performed to evaluate the number of EFA latent factors and the inclusion of the predictor \texttt{age}. In particular, we assessed whether a standard set of IC (i.e., AIC, CAIC, BIC, ssBIC) and CC (i.e., CLC, ICL-BIC, and H) would identify the best classifier (for further details on calculation see the online supplementary material). All the indices agreed on selecting the models with $K=1$ and predictor \texttt{age}, demonstrating their ability to identify the best classifier according to BACC, MCC, SE, and SP, which also reached a satisfactory classification accuracy. By considering the SE metric, the selected model has also shown to correctly identify all the fakers.

\begin{table}[!t]
	\caption{Case study 1: Selection procedure for the best number of EFA factors $K$ and the inclusion of \texttt{age} predictor in the CFA\texttt{+}EFA model, based on IC, CL, and H indices along with the BACC, MCC, SE, and SP metrics of classifications. The best values for each of the selection indices are reported in bold. \label{tab:4}}
	\resizebox{!}{1.7cm}{
		\begin{tabular}{lcccccccccccc}
			\toprule
			Model & $\ell$ & $\text{AIC}$ & $\text{CAIC}$ & $\text{BIC}$ & $\text{ssBIC}$ & $\text{CLC}$ & $\text{ICL-BIC}$ & $\text{H}$ & $\text{BACC}$ & $\text{MCC}$ & $\text{SE}$ & $\text{SP}$\\
			\midrule
			$K=1$ & $-31202.9928$ & $62767.9857$ & $63781.5623$ & $63600.5623$ & $63025.8264$ & $62406.6896$ & $63601.2661$ & $0.9993$ & $0.9541$ & $0.8328$ & $\boldsymbol{1.0}$ & $0.9083$\\
			$K=2$ & $-32976.8191$ & $66401.6381$ & $67656.0091$ & $67432.0091$ & $66720.7338$ & $65955.4852$ & $67433.8562$ & $0.9982$ & $0.8685$ & $0.7069$ & $0.8438$ & $0.8932$\\
			$K=3$ & $-34668.269$ & $69870.538$ & $71365.7034$ & $71098.7034$ & $70250.8886$ & $69365.579$ & $71127.7444$ & $0.9715$ & $0.8932$ & $0.7654$ & $0.8706$ & $0.9157$\\
			$K=4$ & $-34494.7831$ & $69609.5662$ & $71345.526$ & $71035.526$ & $70051.1717$ & $69090.6894$ & $71136.6492$ & $0.9008$ & $0.7057$ & $0.4443$ & $0.4751$ & $0.9363$\\
			$K=5$ & $-34366.113$ & $69438.2261$ & $71414.9804$ & $71061.9804$ & $69941.0866$ & $68761.9446$ & $71091.6989$ & $0.9708$ & $0.8238$ & $0.6625$ & $0.7364$ & $0.9113$\\
			$K=6$ & $-35094.2651$ & $70980.5303$ & $73198.079$ & $72802.079$ & $71544.6458$ & $70262.8641$ & $72876.4128$ & $0.927$ & $0.6133$ & $0.2$ & $0.3459$ & $0.8808$\\
			$K=1 \wedge \texttt{age}$ & $\boldsymbol{-31117.9776}$ & $\boldsymbol{62597.9553}$ & $\boldsymbol{63611.5318}$ & $\boldsymbol{63430.5318}$ & $\boldsymbol{62855.7959}$ & $\boldsymbol{62236.1264}$ & $\boldsymbol{63430.703}$ & $\boldsymbol{0.9998}$ & $\boldsymbol{0.9549}$ & $\boldsymbol{0.836}$ & $\boldsymbol{1.0}$ & $0.9099$\\
			$K=2 \wedge \texttt{age}$ & $-35068.9146$ & $70585.8293$ & $71840.2003$ & $71616.2003$ & $70904.9249$ & $70139.6536$ & $71618.0246$ & $0.9982$ & $0.9218$ & $0.7951$ & $0.9341$ & $0.9096$\\
			$K=3 \wedge \texttt{age}$ & $-32890.4911$ & $66314.9821$ & $67810.1475$ & $67543.1475$ & $66695.3327$ & $65872.8352$ & $67635.0006$ & $0.9099$ & $0.7174$ & $0.4729$ & $0.4975$ & $\boldsymbol{0.9373}$\\
			$K=4 \wedge \texttt{age}$ & $-34406.6492$ & $69433.2984$ & $71169.2582$ & $70859.2582$ & $69874.9039$ & $68832.182$ & $70878.1419$ & $0.9815$ & $0.8982$ & $0.7573$ & $0.893$ & $0.9033$\\
			$K=5 \wedge \texttt{age}$ & $-35102.9369$ & $70911.8737$ & $72888.628$ & $72535.628$ & $71414.7342$ & $70207.9612$ & $72537.7155$ & $0.998$ & $0.9157$ & $0.7948$ & $0.9158$ & $0.9156$\\
			$K=6 \wedge \texttt{age}$ & $-34913.6591$ & $70619.3183$ & $72836.867$ & $72440.867$ & $71183.4338$ & $69845.7082$ & $72459.2569$ & $0.982$ & $0.8989$ & $0.779$ & $0.8768$ & $0.9211$\\
			\bottomrule
	\end{tabular}}	
	
\end{table}

\noindent Table \ref{tab:5} shows the estimated parameters for the selected model along with the bootstrap standard errors (with 1500 replicates). The results of the CFA submodel are in line with the BFI latent structure. In particular, $\hat{\boldsymbol{\Lambda}}_1$ presents mostly observed items with medium-to-strong factor loadings. On the other hand, $\hat{\boldsymbol{\Lambda}}_2$ is characterized by extremely strong factor loadings, possibly due to the inflated inter-item correlations induced by faking. Considering the measurement errors, $\hat{\boldsymbol{\Theta}}_{\boldsymbol{\delta}}$ shows that errors of the CFA component are generally between 0.3 and 0.9, whereas the errors of EFA component are lower and occasionally equal to zero. Considering the correlation matrix of the CFA latent factors $\hat{\boldsymbol{\Phi}}$, the values  mostly indicate small or irrelevant magnitudes of correlation among the latent factors. Finally, the logistic regression parameter $\hat{\boldsymbol{\beta}}$ shows that an increase of one year in \texttt{age} is associated to a decrease of 0.6\% in the relative probability of being classified into the CFA submodel  ($\exp(\hat{\beta}_{\texttt{age}})=0.994,\sigma_{\hat{\boldsymbol{\beta}}}=0.330$).

\begin{table}[!t]
	\caption{Case study 1: Estimates of the selected model CFA\texttt{+}EFA model with  bootstrap standard errors.\label{tab:5}}
	\resizebox{!}{6cm}{
		\begin{tabular}{lcccccccc}
			\toprule
			\multicolumn{6}{c}{$\hat{\boldsymbol{\Lambda}}_1$}& \multirow{2}{*}{$\hat{\boldsymbol{\Theta}}_{\boldsymbol{\delta}}$} & $\hat{\boldsymbol{\Lambda}}_2$ & \multirow{2}{*}{$\hat{\boldsymbol{\Psi}}_{\boldsymbol{\delta}}$}\\\cmidrule{2-6} \cmidrule{8-8}
			&$q=1$ &$q=2$ &$q=3$ &$q=4$ &$q=5$ &&$K=1$&\\
			\midrule	
			\quad$\text{Item 1}$ & $1.033\,(0.666)$ & $0.0$ & $0.0$ & $0.0$ & $0.0$ & $0.0\,(0.321)$ & $-0.602\,(0.71$) & $0.218\,(0.453)$\\
			\quad$\text{Item 2}$ & $0.398\,(0.469)$ & $0.0$ & $0.0$ & $0.0$ & $0.0$ & $0.841\,(0.297)$ & $-0.72\,(0.758)$ & $0.257\,(0.523)$\\
			\quad$\text{Item 3}$ & $0.336\,(0.311)$ & $0.0$ & $0.0$ & $0.0$ & $0.0$ & $0.912\,(0.398)$ & $-0.753\,(0.879)$ & $0.086\,(0.523)$\\
			\quad$\text{Item 4}$ & $-0.444\,(0.558)$ & $0.0$ & $0.0$ & $0.0$ & $0.0$ & $0.769\,(0.51$) & $0.782\,(0.416)$ & $0.29\,(0.482)$\\
			\quad$\text{Item 5}$ & $0.296\,(0.254)$ & $0.0$ & $0.0$ & $0.0$ & $0.0$ & $0.959\,(0.37$) & $-0.768\,(0.747)$ & $0.0\,(0.419)$\\
			\quad$\text{Item 6}$ & $-0.492\,(0.589)$ & $0.0$ & $0.0$ & $0.0$ & $0.0$ & $0.772\,(0.386)$ & $0.608\,(0.549)$ & $0.425\,(0.536)$\\
			\quad$\text{Item 7}$ & $0.589\,(0.503)$ & $0.0$ & $0.0$ & $0.0$ & $0.0$ & $0.755\,(0.163)$ & $-0.412\,(0.529)$ & $0.355\,(0.225)$\\
			\quad$\text{Item 8}$ & $0.176\,(0.489)$ & $0.0$ & $0.0$ & $0.0$ & $0.0$ & $0.668\,(0.303)$ & $1.257\,(0.363)$ & $0.049\,(0.293)$\\
			\quad$\text{Item 9}$ & $0.0$ & $-0.929\,(0.463)$ & $0.0$ & $0.0$ & $0.0$ & $0.3\,(0.453)$ & $-0.67\,(0.813)$ & $0.0\,(0.452)$\\
			\quad$\text{Item 10}$ & $0.0$ & $0.706\,(0.648)$ & $0.0$ & $0.0$ & $0.0$ & $0.639\,(0.278)$ & $0.545\,(0.825)$ & $0.0\,(0.259)$\\
			\quad$\text{Item 11}$ & $0.0$ & $0.479\,(0.546)$ & $0.0$ & $0.0$ & $0.0$ & $0.771\,(0.305)$ & $0.706\,(0.665)$ & $0.242\,(0.153)$\\
			\quad$\text{Item 12}$ & $0.0$ & $0.716\,(0.687)$ & $0.0$ & $0.0$ & $0.0$ & $0.535\,(0.309)$ & $0.762\,(0.75$) & $0.078\,(0.237)$\\
			\quad$\text{Item 13}$ & $0.0$ & $-0.201\,(0.557)$ & $0.0$ & $0.0$ & $0.0$ & $0.775\,(0.238)$ & $1.111\,(0.838)$ & $0.132\,(0.326)$\\
			\quad$\text{Item 14}$ & $0.0$ & $0.372\,(0.357)$ & $0.0$ & $0.0$ & $0.0$ & $0.872\,(0.495)$ & $0.667\,(0.936)$ & $0.259\,(0.844)$\\
			\quad$\text{Item 15}$ & $0.0$ & $0.636\,(0.366)$ & $0.0$ & $0.0$ & $0.0$ & $0.651\,(0.336)$ & $0.656\,(0.377)$ & $0.383\,(0.355)$\\
			\quad$\text{Item 16}$ & $0.0$ & $-0.531\,(0.655)$ & $0.0$ & $0.0$ & $0.0$ & $0.741\,(0.892)$ & $-0.728\,(0.787)$ & $0.143\,(0.787)$\\
			\quad$\text{Item 17}$ & $0.0$ & $-0.328\,(0.635)$ & $0.0$ & $0.0$ & $0.0$ & $0.769\,(0.347)$ & $1.086\,(0.759)$ & $0.035\,(0.285)$\\
			\quad$\text{Item 18}$ & $0.0$ & $0.0$ & $0.764\,(0.459)$ & $0.0$ & $0.0$ & $0.285\,(0.346)$ & $-0.997\,(0.59$) & $0.186\,(0.205)$\\
			\quad$\text{Item 19}$ & $0.0$ & $0.0$ & $-0.449\,(0.629)$ & $0.0$ & $0.0$ & $0.884\,(0.263)$ & $-0.495\,(0.483)$ & $0.378\,(0.481)$\\
			\quad$\text{Item 20}$ & $0.0$ & $0.0$ & $0.31\,(0.385)$ & $0.0$ & $0.0$ & $0.961\,(0.254)$ & $0.208\,(0.461)$ & $0.728\,(0.157)$\\
			\quad$\text{Item 21}$ & $0.0$ & $0.0$ & $-0.351\,(0.499)$ & $0.0$ & $0.0$ & $0.884\,(0.259)$ & $-0.774\,(0.492)$ & $0.225\,(0.242)$\\
			\quad$\text{Item 22}$ & $0.0$ & $0.0$ & $0.541\,(0.428)$ & $0.0$ & $0.0$ & $0.789\,(0.223)$ & $0.514\,(0.365)$ & $0.407\,(0.194)$\\
			\quad$\text{Item 23}$ & $0.0$ & $0.0$ & $0.58\,(0.532)$ & $0.0$ & $0.0$ & $0.403\,(0.37$) & $-1.19\,(0.607)$ & $0.077\,(0.334)$\\
			\quad$\text{Item 24}$ & $0.0$ & $0.0$ & $-0.368\,(0.536)$ & $0.0$ & $0.0$ & $0.918\,(0.339)$ & $-0.691\,(0.639)$ & $0.216\,(0.352)$\\
			\quad$\text{Item 25}$ & $0.0$ & $0.0$ & $0.664\,(0.369)$ & $0.0$ & $0.0$ & $0.508\,(0.249)$ & $-0.842\,(0.408)$ & $0.324\,(0.152)$\\
			\quad$\text{Item 26}$ & $0.0$ & $0.0$ & $0.515\,(0.542)$ & $0.0$ & $0.0$ & $0.423\,(0.296)$ & $-1.259\,(0.698)$ & $0.038\,(0.629)$\\
			\quad$\text{Item 27}$ & $0.0$ & $0.0$ & $0.159\,(0.284)$ & $0.0$ & $0.0$ & $0.917\,(0.316)$ & $0.637\,(0.653)$ & $0.687\,(0.252)$\\
			\quad$\text{Item 28}$ & $0.0$ & $0.0$ & $0.0$ & $-0.911\,(0.399)$ & $0.0$ & $0.367\,(0.595)$ & $-0.508\,(1.165)$ & $0.061\,(0.788)$\\
			\quad$\text{Item 29}$ & $0.0$ & $0.0$ & $0.0$ & $-0.756\,(0.287)$ & $0.0$ & $0.562\,(0.341)$ & $-0.422\,(0.74$) & $0.367\,(0.544)$\\
			\quad$\text{Item 30}$ & $0.0$ & $0.0$ & $0.0$ & $0.421\,(0.234)$ & $0.0$ & $0.842\,(0.231)$ & $0.453\,(0.687)$ & $0.544\,(0.217)$\\
			\quad$\text{Item 31}$ & $0.0$ & $0.0$ & $0.0$ & $-0.542\,(0.314)$ & $0.0$ & $0.709\,(0.376)$ & $-0.547\,(0.75$) & $0.545\,(0.177)$\\
			\quad$\text{Item 32}$ & $0.0$ & $0.0$ & $0.0$ & $-0.393\,(0.36$) & $0.0$ & $0.833\,(0.507)$ & $-0.512\,(0.388)$ & $0.715\,(0.35$)\\
			\quad$\text{Item 33}$ & $0.0$ & $0.0$ & $0.0$ & $-0.261\,(0.437)$ & $0.0$ & $0.788\,(0.464)$ & $0.993\,(0.831)$ & $0.283\,(0.265)$\\
			\quad$\text{Item 34}$ & $0.0$ & $0.0$ & $0.0$ & $-0.274\,(0.394)$ & $0.0$ & $0.768\,(0.433)$ & $0.91\,(0.442)$ & $0.529\,(0.322)$\\
			\quad$\text{Item 35}$ & $0.0$ & $0.0$ & $0.0$ & $0.551\,(0.284)$ & $0.0$ & $0.899\,(0.323)$ & $0.364\,(0.776)$ & $0.065\,(0.308)$\\
			\quad$\text{Item 36}$ & $0.0$ & $0.0$ & $0.0$ & $0.0$ & $-0.188\,(0.363)$ & $1.147\,(0.615)$ & $0.479\,(1.083)$ & $0.0\,(1.174)$\\
			\quad$\text{Item 37}$ & $0.0$ & $0.0$ & $0.0$ & $0.0$ & $0.855\,(0.684)$ & $0.308\,(0.318)$ & $-0.812\,(0.53$) & $0.096\,(0.247)$\\
			\quad$\text{Item 38}$ & $0.0$ & $0.0$ & $0.0$ & $0.0$ & $-0.458\,(0.536)$ & $0.798\,(0.716)$ & $0.815\,(0.573)$ & $0.056\,(0.447)$\\
			\quad$\text{Item 39}$ & $0.0$ & $0.0$ & $0.0$ & $0.0$ & $-0.373\,(0.462)$ & $0.829\,(0.387)$ & $0.601\,(0.313)$ & $0.642\,(0.196)$\\
			\quad$\text{Item 40}$ & $0.0$ & $0.0$ & $0.0$ & $0.0$ & $-0.05\,(0.253)$ & $0.716\,(0.34$) & $1.204\,(0.785)$ & $0.115\,(0.225)$\\
			\quad$\text{Item 41}$ & $0.0$ & $0.0$ & $0.0$ & $0.0$ & $0.824\,(0.664)$ & $0.282\,(0.304)$ & $-0.825\,(0.545)$ & $0.289\,(0.263)$\\
			\quad$\text{Item 42}$ & $0.0$ & $0.0$ & $0.0$ & $0.0$ & $-0.665\,(0.349)$ & $0.449\,(0.292)$ & $0.706\,(0.723)$ & $0.71\,(0.186)$\\
			\midrule
			$\hat{\boldsymbol{\Phi}}$&&&&&&&&\\
			\quad${q=1}$ &  $1.0$&&&&&&&\\
			\quad${q=2}$ & $-0.154\,(0.279)$ & $1.0$ &&&&&&\\
			\quad${q=3}$ & $-0.024\,(0.293)$ & $0.041\,(0.261)$ & $1.0$&&&&&\\
			\quad${q=4}$ & $-0.21\,(0.273)$ & $0.464\,(0.277)$ & $0.155\,(0.28$) & $1.0$&&&&\\
			\quad${q=5}$ & $0.193\,(0.35$) & $-0.352\,(0.36$) & $0.103\,(0.312)$ & $-0.318\,(0.358)$ & $1.0$&&&\\
			$\hat{\boldsymbol{\nu}}$&&&&&&&&\\
			\quad$K=1$ &$-1.119\,(0.746)$&&&&&&&\\
			&&&&&&&&\\
			$\hat{\boldsymbol{\beta}}$&&&&&&&&\\
			$\quad \hat{\beta}_0$ & $1.209$&&&&&&&\\
			\quad$\hat{\beta}_{\texttt{age}}$ &$-0.006\,(0.330)$&&&&&&&\\
			\bottomrule
		\end{tabular}
	}
	\centering
\end{table}

\subsection{Case study 2: Careless responding}\label{sec:5.2}

\paragraph{\textit{Data and measures}}
The dataset consists of $n=708$ participants ($60.09\%$ male, with ages ranging from 18 to 75, a mean age of $34.6$, and a standard deviation of $11.69$) who responded to $p=36$ items drawn from the Big Five (BF; \cite{goldberg1992development}). The dataset is publicly available at \url{https://osf.io/3fw59/}. The items were scored on a 5-point scale ranging from 1 (``very inaccurate") to 5 (``very accurate") which were intended to measure three latent factors: extraversion, emotional stability, and conscientiousness. For each latent factor, six items were used to describe the factor in positive  polarity (e.g., \textquotedblleft Bold\textquotedblright) and six items in negative polarity (e.g., \textquotedblleft Timid\textquotedblright)  \cite{arias2020little}.  The final data matrix corresponds to $\boldsymbol{Y}_{708\times36}$. Unlike for the Case study 1, none aberrant response behaviour has been experimentally induced so that data naturally consist of aberrant as well as non-aberrant responses with no prior knowledge on the latter. However, previous analyses on the same dataset have revealed the presence of careless responding \cite{arias2020little}. Two covariates were also included to predict the mixture parameter, namely \texttt{age} in years and \texttt{gender} (\texttt{m} vs. \texttt{f}).

\paragraph{\textit{Exploratory analyses}}
Following the data analysis workflow described in Case study 1, also in this case a preliminary exploratory analysis has been performed. In particular, to verify the quality of the latent structure present in the dataset, a CFA model based on the BF factor structure (see \cite{goldberg1992development}) has been fit on the observed sample. The results were evaluated in terms of CFI and RMSEA, which showed that the CFA latent structure is not acceptable ($\text{CFI}=0.643$ and $\text{RMSEA}=0.110$), in line with the findings provided by \cite{arias2020little}.

\paragraph{\textit{CFA\texttt{+}EFA analysis}}

The CFA\texttt{+}EFA model has been specified such that the CFA submodel reproduces the BF latent structure. For the sake of simplicity, Table \ref{tab:8} reports a subset of results from the model selection procedure, which was performed to assess the number of EFA latent factors when \texttt{age} and \texttt{RT} are also included (see Table S1 in the online supplementary material for the complete model selection table). We investigated a wide range of $K$ values to disentangle careless responding. Considering the properties of the entropy index, we resorted to select the model with the lowest values of ICs and CCs conditioning on the models with the highest entropy values (i.e., the model $K=17\wedge\texttt{gender}\wedge\texttt{age}$). According to the predicted classifications in the mixture variable $\hat{\boldsymbol{z}}$, the selected model indicates that the $6.90\%$ of participants were adopting a careless response behaviour, which is not too far from the proportion identified by \cite{arias2020little}. 

\addtolength{\tabcolsep}{1pt}   
\begin{table}[!t]
	\caption{Case study 2: Model selection procedure to determine the number of EFA latent factors within the CFA\texttt{+}EFA model and the inclusion of the  predictors \texttt{age} and \texttt{RT} based on IC, CC, and H indices.\label{tab:8}}
	\resizebox{!}{3.85cm}{
		\begin{tabular}{lcccccccc}
			\toprule
			Model & $\ell$ & $\text{AIC}$ & $\text{CAIC}$ & $\text{BIC}$ & $\text{ssBIC}$ & $\text{CLC}$ & $\text{ICL-BIC}$ & $\text{H}$\\
			\midrule
			$K=1$&	$-28877.941$ & $58053.881$ & $58882.685$ & $58733.685$ & $58260.575$ & $57916.328$ & $58894.132$ & $0.837$\\
			$K=5$&	$-31107.771$ & $62809.541$ & $64461.587$ & $64164.587$ & $63221.543$ & $62235.4$ & $64184.446$ & $0.98$\\
			$K=10$&	$-31489.177$ & $63942.355$ & $66623.453$ & $66141.453$ & $64610.99$ & $62979.413$ & $66142.511$ & $0.999$\\
			$K=15$&	$-30216.694$ & $61767.388$ & $65477.538$ & $64810.538$ & $62692.658$ & $60463.08$ & $64840.23$ & $0.97$\\
			$K=20$&	$-30180.78$ & $62065.56$ & $66804.762$ & $65952.762$ & $63247.464$ & $60378.902$ & $65970.105$ & $0.982$\\
			$K=1\wedge\texttt{gender}$&	$-30149.642$ & $60601.283$ & $61441.212$ & $61290.212$ & $60810.752$ & $60454.703$ & $61445.632$ & $0.842$\\
			$K=5\wedge\texttt{gender}$&	$-28580.055$ & $57758.11$ & $59421.28$ & $59122.28$ & $58172.886$ & $57285.685$ & $59247.855$ & $0.872$\\
			$K=10\wedge\texttt{gender}$&	$-30869.652$ & $62707.305$ & $65399.528$ & $64915.528$ & $63378.715$ & $61815.534$ & $64991.757$ & $0.922$\\
			$K=15\wedge\texttt{gender}$&	$-30171.913$ & $61681.826$ & $65403.102$ & $64734.102$ & $62609.871$ & $60353.928$ & $64744.203$ & $0.99$\\
			$K=20\wedge\texttt{gender}$&	$-29977.138$ & $61662.276$ & $66412.603$ & $65558.603$ & $62846.954$ & $59974.439$ & $65578.767$ & $0.979$\\
			$K=1\wedge\texttt{age}$&	$-29696.995$ & $59693.991$ & $60528.357$ & $60378.357$ & $59902.072$ & $59549.561$ & $60533.928$ & $0.841$\\
			$K=5\wedge\texttt{age}$&	$-30240.527$ & $61077.054$ & $62734.662$ & $62436.662$ & $61490.443$ & $60544.01$ & $62499.619$ & $0.936$\\
			$K=10\wedge\texttt{age}$&	$-30588.455$ & $62142.91$ & $64829.57$ & $64346.57$ & $62812.933$ & $61176.91$ & $64346.57$ & $1.0$\\
			$K=15\wedge\texttt{age}$&	$-30189.018$ & $61714.037$ & $65429.749$ & $64761.749$ & $62640.694$ & $60382.483$ & $64766.196$ & $0.995$\\
			$K=20\wedge\texttt{age}$&	$-29512.164$ & $60730.328$ & $65475.093$ & $64622.093$ & $61913.619$ & $59025.395$ & $64623.16$ & $0.999$\\
			$K=1\wedge\texttt{gender}\wedge\texttt{age}$&$-30500.882$ & $61309.764$ & $62166.38$ & $62012.38$ & $61523.395$ & $61156.748$ & $62167.365$ & $0.842$\\
			$K=5\wedge\texttt{gender}\wedge\texttt{age}$&	$-28988.196$ & $58580.392$ & $60260.250$ & $59958.250$ & $58999.329$ & $58094.187$ & $60076.045$ & $0.880$\\
			$K=10\wedge\texttt{gender}\wedge\texttt{age}$&	$-30624.047$ & $62222.094$ & $64931.004$ & $64444.004$ & $62897.666$ & $61251.174$ & $64447.084$ & $0.997$\\
			$K=15\wedge\texttt{gender}\wedge\texttt{age}$&	$-30319.927$ & $61983.855$ & $65721.817$ & $65049.817$ & $62916.06$ & $60655.184$ & $65065.147$ & $0.984$\\
			$K=17\wedge\texttt{gender}\wedge\texttt{age}$&	$-29541.254$ & $60574.508$ & $64724.091$ & $63978.091$ & $61609.367$ & $59082.508$ & $63978.091$ & $1.0$\\
			$K=20\wedge\texttt{gender}\wedge\texttt{age}$&	$-29655.578$ & $61025.157$ & $65792.171$ & $64935.171$ & $62213.997$ & $59314.66$ & $64938.675$ & $0.996$\\
			\bottomrule
	\end{tabular}}
	\centering
\end{table}
\addtolength{\tabcolsep}{-1pt}

\begin{figure}[!t]
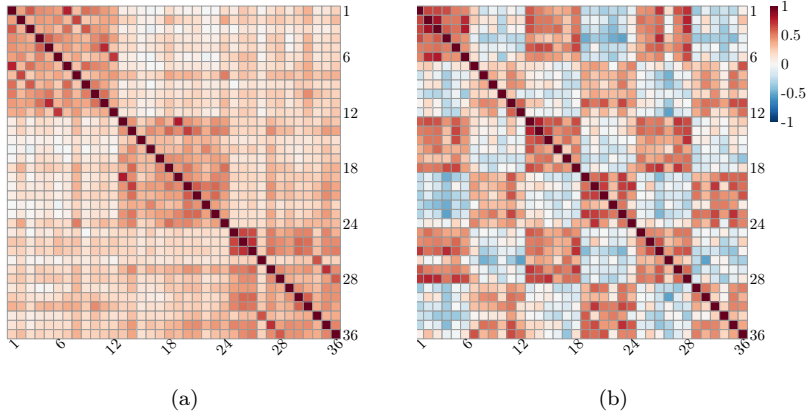

	\centering
	\hspace{-0.3cm}
	\subfloat[][]{%
		\resizebox{5.2cm}{!}{\input{fig_carefull_heat.tex}}\label{carf_heat}%
	}\hspace{0.5cm}%
	\subfloat[][]{%
		\resizebox{5.2cm}{!}{\input{fig_careless_heat.tex}}\label{carl_heat}%
	}
	\caption{Case study 2: Heatmaps of the predicted inter-item correlation matrices. Panel (a) represents the predicted careful subsample's correlation matrix (CFA component); Panel (b) the predicted careless subsample's correlation matrix (EFA component).}
	\label{fig_heat_careless}
\end{figure}

Figure \ref{fig_heat_careless} shows the heatmaps of inter-item correlations in the predicted subsample of careful and careless respondents. Figure \ref{carf_heat} presents a correlation matrix in which the three factor latent structure of the BF items (i.e., correlation blocks along the main diagonal) can be distinguished, even in presence of inflated correlations outside these blocks. Figure \ref{carl_heat} reports an unexpected pattern of correlation between the BF items, which indicates that, within each factor, the  positive-polarity items are slightly negatively correlated with the  negative-polarity items for the same factor. In other words, the respondents used (carelessly) the 5-point scale to respond to the negative-polarity items as they were responding to a positive-polarity item. For instance, a careless respondent answered ``very accurate" to the ``Bold" items (positive polarity), but answered ``very accurate" to the ``Timid" items as well (negative polarity). On the other hand, the correlations among the factors indicates that these respondents used the scale with a homogeneous careless behaviour across the items.

\begin{figure}[!t]
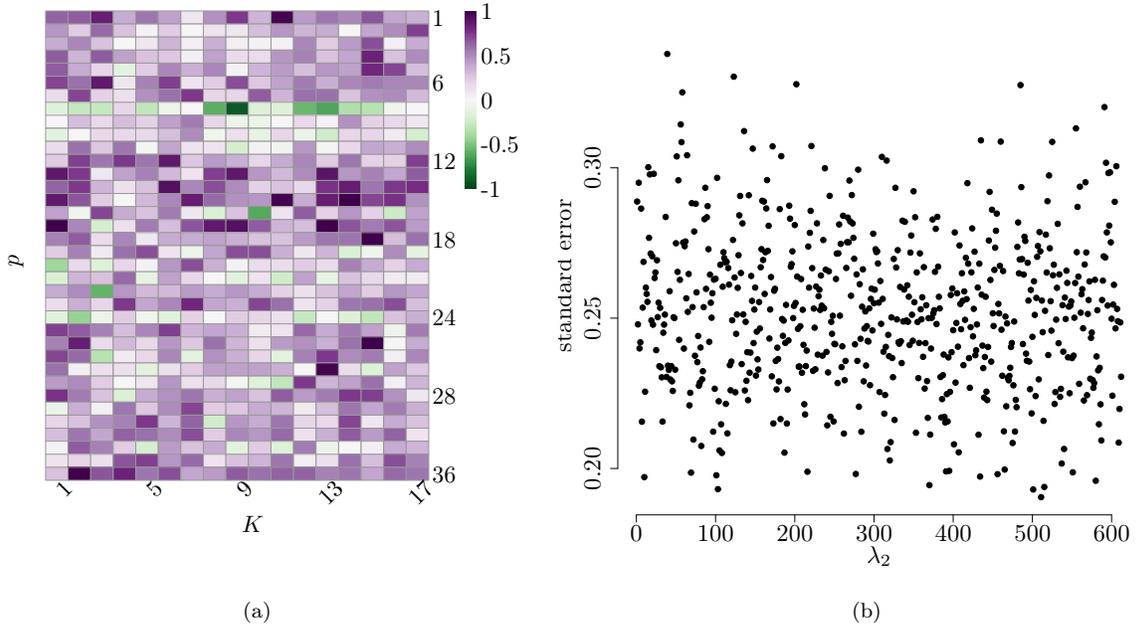

	\hspace{-0.8cm}
	\subfloat[][]{%
		\resizebox{8cm}{!}{\input{lambdas.tex}}\label{lamb}%
	}\hspace{0cm}%
	\subfloat[][]{%
		\resizebox{8.2cm}{!}{\input{standerr.tex}}\label{se}%
	}
	\caption{Case study 2: Graphic representation of $\hat{\boldsymbol{\Lambda}}_2$ and its bootstrap standard errors. Panel (a) heatmap of $\hat{\boldsymbol{\Lambda}}_2$; Panel (b) bootstrap standard errors of $\hat{\boldsymbol{\Lambda}}_2$.}
	\label{L2}
\end{figure}

\begin{figure}[!t]
	\centering
	
	\resizebox{6cm}{!}{\input{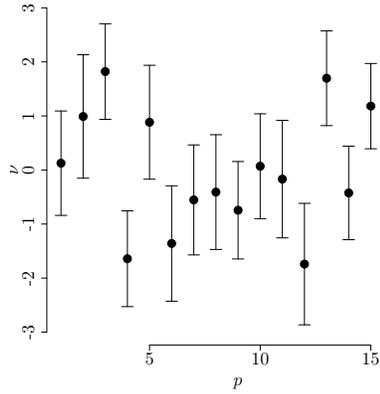}}
	\caption{Case study 2: Graphic representation of $\hat{\boldsymbol{\nu}}$ with its bootstrap standard errors.}
	\label{nu}
\end{figure}

Table \ref{tab:9} shows the estimated parameters of the model with $K=17$ latent factor and the predictors \texttt{gender} and \texttt{age} (bootstrap standard errors were calculated with 1500 bootstrap samples). By examining the parameter estimates, we observe that $\hat{\boldsymbol{\Lambda}}_1$ shows loadings generally above 0.40, which suggest a medium-to-large association between latent and observed variables. Considering the measurement errors, $\hat{\boldsymbol{\Psi}}_{\boldsymbol{\epsilon}}$ shows that the variances of the EFA errors are extremely lower than those of the CFA model. With respect to the logistic regression parameter $\hat{\boldsymbol{\beta}}$, the estimate indicates that being an \texttt{f} vs. an \texttt{m} participant is associated to an increase of 653.8\% in the relative probability of being assigned to the CFA submodel   ($\exp(\hat{\beta}_{\texttt{gender}})=7.538,\sigma_{\hat{\boldsymbol{\beta}}}=0.603$), whereas an increase of one year in \texttt{age} is related to a decrease of 2.3\% in the relative probability of being assigned to the CFA submodel ($\exp(\hat{\beta}_{\texttt{age}})=0.972,\sigma_{\hat{\boldsymbol{\beta}}}=0.021$). To complete the results presentation, Figures \ref{L2} and \ref{nu} show the estimates for the EFA component with $K=17$ latent variables in a graphical format. They highlight that $\hat{\boldsymbol{\Lambda}}_2$ has parameters of varying magnitude between latent factors and observed indicators, although the majority of them are generally positive.

\addtolength{\tabcolsep}{4pt} 
\begin{table}[!t]
	\caption{Case study 2: Estimates of the CFA\texttt{+}EFA model with  bootstrap standard errors. 	\label{tab:9}}
	\resizebox{!}{7.8cm}{
		\begin{tabular}{lccccc}
			\toprule
			&\multicolumn{3}{c}{$\hat{\boldsymbol{\Lambda}}_1$}& \multirow{2}{*}{$\hat{\boldsymbol{\Theta}}_{\boldsymbol{\delta}}$} & \multirow{2}{*}{$\hat{\boldsymbol{\Psi}}_{\boldsymbol{\delta}}$}\\\cmidrule{2-4} 
			&$q=1$ &$q=2$ &$q=3$ & &\\
			\midrule	
			\quad$\text{Item 1}$ & $0.812\,(0.122)$ & $0.0$ & $0.0$ & $0.312\,(0.127)$ & $0.025\,(0.062)$\\
			\quad$\text{Item 2}$ & $0.495\,(0.122)$ & $0.0$ & $0.0$ & $0.7\,(0.136)$ & $0.0\,(0.054$)\\
			\quad$\text{Item 3}$ & $0.75\,(0.113)$ & $0.0$ & $0.0$ & $0.458\,(0.121)$ & $0.066\,(0.071)$\\
			\quad$\text{Item 4}$ & $0.613\,(0.137)$ & $0.0$ & $0.0$ & $0.582\,(0.101)$ & $0.015\,(0.077)$\\
			\quad$\text{Item 5}$ & $0.648\,(0.146)$ & $0.0$ & $0.0$ & $0.553\,(0.119)$ & $0.003\,(0.042)$\\
			\quad$\text{Item 6}$ & $0.494\,(0.124)$ & $0.0$ & $0.0$ & $0.713\,(0.163)$ & $0.048\,(0.063)$\\
			\quad$\text{Item 7}$ & $0.766\,(0.151)$ & $0.0$ & $0.0$ & $0.407\,(0.145)$ & $0.0\,(0.086$)\\
			\quad$\text{Item 8}$ & $0.366\,(0.13$) & $0.0$ & $0.0$ & $0.792\,(0.172)$ & $0.229\,(0.120)$\\
			\quad$\text{Item 9}$ & $0.682\,(0.12$) & $0.0$ & $0.0$ & $0.536\,(0.127)$ & $0.023\,(0.062)$\\
			\quad$\text{Item 10}$ & $0.596\,(0.129)$ & $0.0$ & $0.0$ & $0.596\,(0.096)$ & $0.039\,(0.065)$\\
			\quad$\text{Item 11}$ & $0.635\,(0.153)$ & $0.0$ & $0.0$ & $0.551\,(0.146)$ & $0.017\,(0.056$)\\
			\quad$\text{Item 12}$ & $0.433\,(0.102)$ & $0.0$ & $0.0$ & $0.745\,(0.15$) & $0.038\,(0.070$)\\
			\quad$\text{Item 13}$ & $0.0$ & $0.585\,(0.223)$ & $0.0$ & $0.631\,(0.185)$ & $0.08\,(0.052)$\\
			\quad$\text{Item 14}$ & $0.0$ & $0.736\,(0.229)$ & $0.0$ & $0.43\,(0.184)$ & $0.0\,(0.092)$\\
			\quad$\text{Item 15}$ & $0.0$ & $0.433\,(0.128)$ & $0.0$ & $0.776\,(0.284)$ & $0.0\,(0.146)$\\
			\quad$\text{Item 16}$ & $0.0$ & $0.53\,(0.199)$ & $0.0$ & $0.675\,(0.184)$ & $0.0\,(0.131)$\\
			\quad$\text{Item 17}$ & $0.0$ & $0.578\,(0.205)$ & $0.0$ & $0.627\,(0.149)$ & $0.0\,(0.084)$\\
			\quad$\text{Item 18}$ & $0.0$ & $0.606\,(0.202)$ & $0.0$ & $0.603\,(0.245)$ & $0.138\,(0.085)$\\
			\quad$\text{Item 19}$ & $0.0$ & $0.659\,(0.234)$ & $0.0$ & $0.514\,(0.131)$ & $0.0\,(0.063)$\\
			\quad$\text{Item 20}$ & $0.0$ & $0.806\,(0.223)$ & $0.0$ & $0.269\,(0.095)$ & $0.059\,(0.072)$\\
			\quad$\text{Item 21}$ & $0.0$ & $0.662\,(0.169)$ & $0.0$ & $0.447\,(0.155)$ & $0.109\,(0.082)$\\
			\quad$\text{Item 22}$ & $0.0$ & $0.546\,(0.148)$ & $0.0$ & $0.563\,(0.163)$ & $0.0\,(0.060)$\\
			\quad$\text{Item 23}$ & $0.0$ & $0.699\,(0.184)$ & $0.0$ & $0.397\,(0.137)$ & $0.0\,(0.066)$\\
			\quad$\text{Item 24}$ & $0.0$ & $0.578\,(0.182)$ & $0.0$ & $0.6\,(0.139)$ & $0.222\,(0.070$)\\
			\quad$\text{Item 25}$ & $0.0$ & $0.0$ & $0.825\,(0.128)$ & $0.374\,(0.079)$ & $0.0\,(0.095)$\\
			\quad$\text{Item 26}$ & $0.0$ & $0.0$ & $0.888\,(0.121)$ & $0.243\,(0.081)$ & $0.003\,(0.048)$\\
			\quad$\text{Item 27}$ & $0.0$ & $0.0$ & $0.88\,(0.12$) & $0.243\,(0.093)$ & $0.0\,(0.051)$\\
			\quad$\text{Item 28}$ & $0.0$ & $0.0$ & $0.308\,(0.168)$ & $0.883\,(0.184)$ & $0.103\,(0.115)$\\
			\quad$\text{Item 29}$ & $0.0$ & $0.0$ & $0.53\,(0.122)$ & $0.635\,(0.18$) & $0.061\,(0.070)$\\
			\quad$\text{Item 30}$ & $0.0$ & $0.0$ & $0.466\,(0.148)$ & $0.703\,(0.199)$ & $0.0\,(0.065)$\\
			\quad$\text{Item 31}$ & $0.0$ & $0.0$ & $0.459\,(0.143)$ & $0.674\,(0.134)$ & $0.015\,(0.074)$\\
			\quad$\text{Item 32}$ & $0.0$ & $0.0$ & $0.638\,(0.145)$ & $0.557\,(0.111)$ & $0.0\,(0.081)$\\
			\quad$\text{Item 33}$ & $0.0$ & $0.0$ & $0.539\,(0.206)$ & $0.626\,(0.1$ )& $0.0\,(0.056)$\\
			\quad$\text{Item 34}$ & $0.0$ & $0.0$ & $0.289\,(0.168)$ & $0.838\,(0.176)$ & $0.074\,(0.057)$\\
			\quad$\text{Item 35}$ & $0.0$ & $0.0$ & $0.457\,(0.115)$ & $0.637\,(0.214)$ & $0.025\,(0.065)$\\
			\quad$\text{Item 36}$ & $0.0$ & $0.0$ & $0.448\,(0.111)$ & $0.696\,(0.141)$ & $0.0\,(0.081$)\\
			\midrule
			$\hat{\boldsymbol{\Phi}}$&&&&&\\
			\quad${q=1}$ & $1.0$ &  &  & & \\
			\quad${q=2}$ & $0.263\,(0.129)$ & $1.0$ &  & &\\
			\quad${q=3}$ & $0.396\,(0.129)$ & $0.455\,(0.214)$ & $1.0$ & &\\
			&&&&&\\
			$\hat{\boldsymbol{\mu}}$&&&&&\\
			&$-0.095\,(0.260)$ & $0.043\,(0.307)$ & $0.002\,(0.228)$&&\\
			&&&&&\\
			$\hat{\boldsymbol{\beta}}$&&&&&\\
			$\quad \hat{\beta}_0$ & $3.001$&&&&\\
			\quad\texttt{gender} &$2.020\,(0.603)$&&&&\\
			\quad\texttt{age} &$-0.028\,(0.021)$&&&&\\
			\bottomrule
		\end{tabular}
	}
	\centering
\end{table}
\addtolength{\tabcolsep}{-4pt}

All in all, if compared to standard CFA analyses, the results highlight that using a CFA\texttt{+}EFA analysis to investigate the latent structure of the BF items would be of greater utility, especially when researchers think that their data are contaminated by aberrant behaviours such as that of careless responding. In particular, this application demonstrates that the proposed method unfolds the (unpredictable) nature of the inter-item correlations induced by careless responding.

\section{Discussion and Conclusions}\label{sec:6}

In this paper, we have proposed a novel method to detect aberrant behaviours. If one considers that some aberrant response styles - such as faking or careless responding - result in a deterioration of the correlation matrix, then statistical models based on covariance analysis (e.g., CFA, SEMs) need to be appropriately generalized in order to cope with this issue. To this end, a mixture of CFA and EFA has been introduced so that the CFA submodel can still be used to represent the latent structure of a questionnaire or survey, while letting the EFA submodel to reproduce the biased inter-item correlations through its set of parameters. Simulations and case studies have been adopted to assess the classification performances and evaluate the properties of the proposed CFA\texttt{+}EFA model. In particular, Simulation study 1 has been designed to study the classification performance of the model under a quite general data perturbation. Instead, Simulation study 2 has been run so as to provide an external validation to the model in the context of faking perturbation. Overall, the results showed to what extent the CFA\texttt{+}EFA model can be used to detect and classify observations affected by aberrant response styles, by  reconstructing - almost optimally - the original data matrix. These findings have been also corroborated by two case studies involving two peculiar types of aberrant behaviours, namely faking and careless responding. {In line with the theoretical/empirical considerations, simulations and case studies also suggested the fitting of a CFA\texttt{+}EFA model with $K=1$ for faking and a more extensive selection procedure for careless responding}. Furthermore, it is noteworthy to point out that, based on the results yielded by simulation and case studies, the EFA submodel does not generally subsume the CFA. This fact allows the EFA submodel to capture the part of observed variability which cannot be absorbed by the CFA submodel. 

One key advantage of the CFA\texttt{+}EFA model lies in its simplicity and flexibility to address aberrant responding behaviours that might affect the inter-item correlations. To our knowledge most FMMs for \textit{post-sampling} heterogeneity have ben designed to handle a single type of aberrant behaviour at a time, limiting their applicability in unsupervised and exploratory contexts. On the contrary, the CFA\texttt{+}EFA model does not necessitate the inclusion of survey-based indicators (i.e., negatively worded items, social desirability items) nor rely exclusively on external variables (i.e., covariates). In addition, unlike for other FMMs, the proposed model has been validated in controlled scenarios, where aberrant response behaviours have been specifically manipulated in order to control for the classification performance. 

However, the proposed model is not exempt from limitations. For instance, some respondents may engage an optimal response strategy for a portion of items before adopting an aberrant response style. In this case, the CFA\texttt{+}EFA model cannot disentangle among varying aberrant behaviours within the same survey. Similarly, our model might suffer from the inability to distinguish among several multiple aberrant response styles provided by the same respondents in the same time. Although this pattern could be recovered in principle - for instance, by including additional covariates to control the mixture parameter - it should be said that the CFA\texttt{+}EFA model has not been specifically developed to handle this type of problem. Another limitation of the current implementation is related to the independence between CFA and EFA submodels. Although this constraint is quite reasonable across several empirical scenarios, there could have been some cases in which parameters of both models need to be not independent (e.g., $\mathbb C\text{ov}[\boldsymbol{\Theta_\delta},\boldsymbol{\Psi_\epsilon}] \neq \mathbf 0$). Finally, another issue which deserves further attention is that of sample size. Indeed, it needs to be sufficiently large (i) to provide reliable estimates for both submodels and (ii) to perform an extensive model selection procedure to find the optimal number of EFA latent factors. This is of particular importance in the case of careless responding, where the proportion of careless against careful observations is generally small.

Further investigations can enhance the findings of this research study. {For instance, the CFA\texttt{+}EFA model could be extended in order to model also dichotomous and ordinal data, adopting the underlying variable approach or the IRT approach for the specification of the two factor analysis components (e.g., see \cite{joreskog2001factor}).} Moreover,
a theoretically-oriented model - e.g., by adopting a copula-based representation to model the aberrant dependencies among the items \cite{krupskii2013factor} - could be used instead of the simple EFA submodel. This could enhance the ability of the proposed method to detect aberrant response styles. Further generalizations might also include submodels which are not strictly based on the Normality assumption for the latent variables \cite{murray2013bayesian}. This would improve the applicability of the proposed model to disentangle between aberrant and non-aberrant responses in general contexts. 

\vspace{1cm}
\subsubsection*{Acknowledgments}
We acknowledge the CINECA award under the ISCRA initiative, for the availability of high performance computing resources and support. 

\subsubsection*{Data availability statement}
The data that support the findings of this work are openly available in Open Science Framework  at \url{https://osf.io/wdv25/} for Case study 1 and at \url{https://osf.io/3fw59/} for Case study 2.  Additionally, the repository containing the publicly available \texttt{Julia} code can be accessed at the link: \url{https://github.com/niccolocao/CFAmixEFA}.
\nocite{arias2020little,krammerrep}

\clearpage
\bibliographystyle{plain}
\bibliography{biblio}

\clearpage
\includepdf[pages=-]{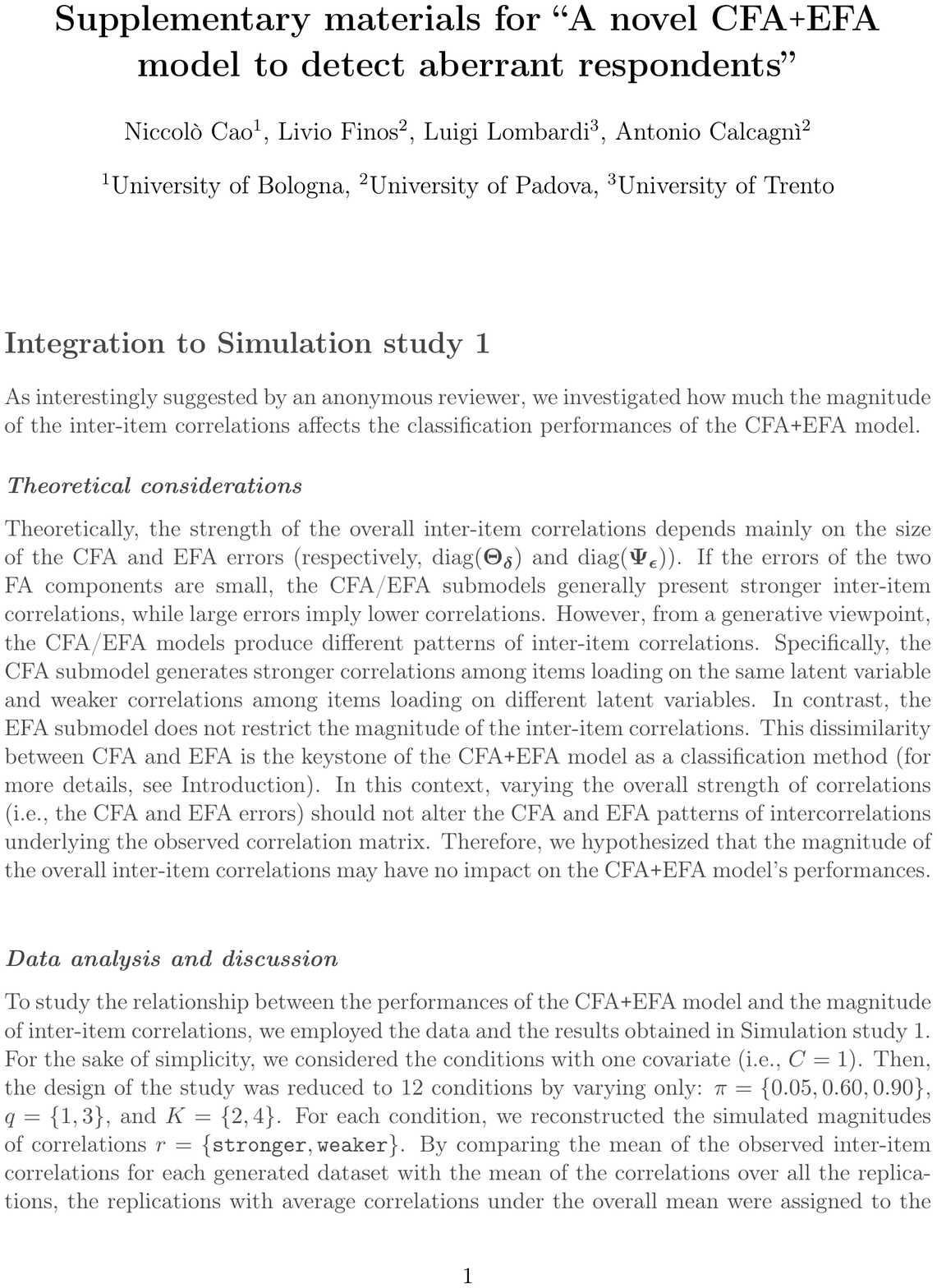}

\end{document}